\documentclass[letter,10pt]{article}
\usepackage{natbib}
\usepackage{amsmath}
\usepackage{amssymb}
\usepackage{multicol}

\usepackage[pdftex]{graphicx}
\usepackage{epstopdf}
\usepackage{latexsym}
\newcounter{Sectioncounter}
\newcounter{SubSectioncounter}

\newcommand{\Section}[1]
{ {
\vspace{20pt}
\addtocounter{Sectioncounter}{1}
\setcounter{SubSectioncounter}{0}
\begin{center}
{\bf \arabic{Sectioncounter}. #1}
\end{center}
} }
\newcommand{\Subsection}[1]
{ {
\vspace{10pt}
\addtocounter{SubSectioncounter}{1}
\begin{center}
{\sc \arabic{Sectioncounter}.\arabic{SubSectioncounter}. \lowercase{#1}}
\end{center}
} }
\newcommand{\Appendix}[1]
{ {
\vspace{20pt}
\begin{center}
{\bf Appendix}
\end{center}
} }

\makeatletter

\newenvironment{figurehere}
  {\def\@captype{figure}}
  {}
\makeatother

\renewcommand{\baselinestretch}{1}

\newcommand{\be}{\begin{equation}}
\newcommand{\ee}{\end{equation}}

\newcommand{\pt}{\partial_t}

\begin{document}

\begin{center}
  {\large \bf Epidemic modeling in metapopulation systems with
  heterogeneous coupling pattern: theory and simulations }\\
  
  \vspace*{0.5cm} 
  
  {\small \sc Vittoria Colizza$^{1,2}$  and Alessandro Vespignani$^{1,2}$}

  \vspace*{0.5cm}

  {\em $^1$ School of Informatics and 
  Biocomplexity Institute, Indiana University, Bloomington IN 47408
  USA\\$^2$ Complex Networks Lagrange Laboratory (CNLL), Institute for
  Scientific Interchange (ISI). Torino, Italy\\
    
    \vspace*{0.5cm} \rm (\today)}

\end{center}

\begin{quotation}
The spatial structure of populations is a key element in the
understanding of the large scale spreading of epidemics. Motivated by
the recent empirical evidence on the heterogeneous properties of
transportation and commuting patterns among urban areas, we present a
thorough analysis of the behavior of infectious diseases in
metapopulation models characterized by heterogeneous connectivity and
mobility patterns. We derive the basic reaction-diffusion equation 
describing the metapopulation system at the mechanistic
level and derive an early stage dynamics approximation for the
subpopulation invasion dynamics. The analytical description uses
degree block variables that allows us to take into account arbitrary
degree distribution of the metapopulation network. We show that along with
the usual single population epidemic threshold the metapopulation
network exhibits a global threshold for the subpopulation
invasion. We find an explicit analytic expression for the invasion
threshold that determines the minimum number of individuals traveling 
among subpopulations in order to have the infection of a macroscopic 
number of subpopulations. The invasion
threshold is a function of factors such as the basic reproductive
number, the infectious period and the mobility process and 
it is found to decrease for increasing network heterogeneity. 
We provide extensive mechanistic numerical Monte Carlo simulations 
that recover the analytical finding in a wide
range of metapopulation network connectivity patterns. The results can
be useful in the understanding of recent data driven computational
approaches to disease spreading  in large transportation
networks and the effect of containment measures such as travel restrictions. 
\end{quotation}

  
\vspace*{0.25cm}

\begin{multicols}{2}

%*****************************************************************************
%*****************************************************************************
\Section{Introduction}
The metapopulation modeling approach is an essential theoretical 
framework used in population ecology, genetics and adaptive evolution
to describe population dynamics whenever the spatial structure
of populations is known to play a key role in the system's evolution~\citep{Hanskibook1,Hanskibook2,Tilman:1997,Bascompte:1998}.
Metapopulation models rely on the basic assumption that the system under study
is characterized by a highly fragmented environment in which the population
is structured and localized in relatively isolated discrete
{\em patches} or subpopulations connected by some degree of migration. 
Classic metapopulation dynamics focuses on the processes of local extinction,
recolonization and regional persistence~\citep{Levins:1969,Levins:1970}, 
as the outcome of the interplay 
between migration processes among unstable local populations and 
population dynamics (e.g birth and death rates, competition and predations). 
This paradigm is  extremely useful also in the case of 
infectious diseases, and can be applied to understand the epidemic dynamics
of spatially structured populations with well defined social units (e.g.
families, villages, city locations, towns, cities, regions) connected
through individual mobility~\citep{Hethcote:1978,May:1979,Anderson:1984,May:1984,Bolker:1993,Bolker:1995,Keeling:2002,Lloyd:1996,Grenfell:1997,Grenfell:1998,Ferguson:2003,Riley:2007}.
The arrival of the infection in any subpopulation and its
epidemic evolution are determined by the
coupling generated by the mobility processes
among subpopulations. The
metapopulation dynamics of infectious diseases has generated a wealth
of models and results considering both mechanistic approaches
taking explicitly into account the movement of individuals~\citep{Baroyan:1969,
Rvachev:1985,Longini:1988,Flahault:1991,sattenspiel,Keeling:2002,Grais:2003} 
and effective coupling
approaches where the diffusion process is expressed as a force of
infection coupling different subpopulations~\citep{Bolker:1995,Lloyd:1996,Earn:1998,Rohani:1999,Keeling:2000,Park:2002,Vazquez:2007}. 
Recently, the
metapopulation approach is being revamped in  computational
approaches for the large scale forecast of infectious disease
spreading~\citep{Grais:2004,Hufnagel:2004,Colizza:2006a,Cooper:2006,Colizza:2007a,Hollingsworth:2006,Riley:2007}. 
 
Metapopulation epidemic models, especially at the mechanistic level, 
are based on the spatial structure of
the environment, and the detailed knowledge of transportation
infrastructures and movement patterns. The increasing computational
power and informatics advances are beginning to lift the 
constraints limiting the collection of large spatiotemporal data on
human behavior and demography, finally allowing for the formulation of
realistic data driven models. On the other hand, 
networks which trace the activities
and interactions of individuals, social patterns, transportation
fluxes, and population movements on a local and global
scale~\citep{amaral01,schnee04,Barrat:2004,Guimera:2005,Chowell:2003}
have been analyzed and found to exhibit complex features encoded in
large scale heterogeneity, self-organization and other properties
typical of complex
systems~\citep{barabasi02,mendesbook,Newman:2003,romuvespibook}. In
particular,  a wide range of societal and
technological networks exhibits  very heterogeneous topologies. The
airport network among cities~\citep{Barrat:2004,Guimera:2005}, the
commuting patterns in inter and intra-urban
areas~\citep{Chowell:2003,transims,DeMontis:2007}, and several
info-structures~\citep{romuvespibook} are indeed characterized by networks
whose nodes, representing the elements of the system, have a wildly
varying degree, i.e. the number of connections to other elements.
These topological fluctuations are mathematically encoded in a
heavy-tailed degree distribution $P(k)$, defined as the probability
that any given node has degree $k$, and have been found to have 
a large impact on epidemic phenomena on complex contact 
patterns~\citep{anderson92,pv01a,pv01b,moreno02,lloyd01,Barthelemy:2005}.

Motivated by the above findings we provide here the analysis of the
behavior of epidemic models in metapopulation systems with
heterogeneous connectivity patterns. In order to have a mechanistic
description of the system, we derive the deterministic
reaction-diffusion equations describing the evolution of the epidemic
in the metapopulation systems. The
heterogeneity of the network is taken explicitly into account by introducing
 degree block variables that
provide results expressed as  functions of the moments of the degree
distribution of the substrate networks. In order to account for
the discreteness of the system and microscopic fluctuations in the
diffusion processes we derive also coarse grained equations for the
 invasion dynamics at the subpopulation level. The system is
characterized by the standard (i.e. single population) epidemic threshold
and by a global invasion threshold providing the condition for the infection 
of a macroscopic number of
subpopulations. The first threshold defines
the usual reproductive number $R_0>1$ that is just
a function of the disease parameters while the second threshold
defines a subpopulations reproductive number $R_*>1$ that 
depends also on the diffusion rate of individuals among
subpopulations~\citep{Ball:1997,Cross:2005,Cross:2007}. 
We find an explicit analytic expression in the limit
$R_0\gtrapprox 1$ for the invasion threshold that is found to depend also
on the network heterogeneity. The larger is the network heterogeneity
and the smaller is the diffusion rate 
that guarantees the invasion of a finite fraction of subpopulations.   
This result provides a framework for the understanding of
the evidence collected on the interplay between travel and  global spread
of infectious diseases~\citep{Viboud:2006} and the poor effectiveness of travel
restrictions in the containment of epidemics~\citep{Cooper:2006,Hollingsworth:2006,Colizza:2007a}. 
Finally, the analytic results are confirmed by mechanistic Monte
Carlo simulations for the infection dynamics in the metapopulation
system, in which each single individual is tracked in time to account 
for the discreteness of the processes involved. Heterogeneous
connectivity patterns among subpopulations are modeled and different
values of the parameters involved are considered to validate the theoretical
results.

The paper is organized as follows. Section 2 introduces
the metapopulation epidemic model on a heterogeneous network of connections
among subpopulations. Two different kinds of mobility processes
are introduced in Section 3 to analyze the stationary
diffusion properties of the system. Section 4 incorporates 
the mobility
processes analyzed in the previous section into a metapopulation epidemic
model. Stochastic effects and
discrete description of the processes are considered
with a tree-like approximation for the analysis of the invasion
dynamics at the level of the subpopulations. The effect of diffusion properties
on the invasion dynamics are analyzed and related to the existence
of an invasion epidemic threshold for the metapopulation system. In Section 5 
the behavior of the system above the invasion threshold  
is studied by mechanistic reaction-diffusion equations 
using a deterministic degree block variables representation.
Finally, in Section 6 we report extensive mechanistic
Monte Carlo simulations which confirm the analytical findings of the previous
sections.

%*****************************************************************************
%*****************************************************************************
\Section{Metapopulation mechanistic model as a microscopic 
reaction-diffusion process}\label{metadef}

Metapopulation models describe spatially structured interacting  subpopulations, such as city locations,
urban areas, or defined geographical regions~\citep{Hanskibook2,Grenfell:1997}. 
Individuals within each subpopulation are divided into
 classes denoting their state
with respect to the modeled disease~\citep{anderson92}---such as infected, susceptible,
immune, etc.---and the compartment dynamics accounts for the possibility
that individuals in the same location may get into contact and change
their state according to the infection dynamics.
The interaction among subpopulations is the result of the movement of
individuals from one subpopulation to the other. 
It is clear that the key issue in such a modeling approach is how 
accurately we can describe the commuting patterns or traveling of
people. In many instances even complicate mechanistic 
 patterns can be accounted for by effective couplings expressed
as a force of infection generated by the infectious individuals in
subpopulation $j$ on the individuals in subpopulation
$i$~\citep{Bolker:1995,Lloyd:1996,Earn:1998,Rohani:1999,Keeling:2000,Park:2002}. 
More realistic descriptions are provided by explicit mechanistic
approaches which include the detailed rate of traveling/commuting
obtained from data or from empirical fit to gravity law models (for a recent
reference, see~\cite{Viboud:2006}), accompanied by 
 the associated mixing subpopulations $N_{ij}$ denoting the number of
individuals of the subpopulation $i$ present in the subpopulation
$j$~\citep{Keeling:2002,sattenspiel}. 

A simplified mechanistic approach uses a markovian assumption 
in which at each time step the
movement of individuals is given according to a matrix $d_{ij}$ that
expresses the probability that an individual in the subpopulation $i$
is traveling to the subpopulation $j$. The markovian character is in
the fact that we do not label individuals according to their original
subpopulation (e.g. \emph{home} in a commuting pattern framework) 
and at each time step the same traveling probability
applies to all individuals in the subpopulation without having memory
of their origin. This approach is extensively used for very large
populations in the case the traffic $w_{ij}$ among
subpopulations is known by stating that $d_{ij}\sim w_{ij}/N_j$. 
Several modeling approaches to
the large scale spreading of infectious disease~\citep{Baroyan:1969,
Rvachev:1985,Longini:1988,Flahault:1991,Grais:2003,Grais:2004,
Hufnagel:2004,Colizza:2006a,Colizza:2006b,Colizza:2007a}
use this mobility process based
on transportation networks for which it is now possible to 
obtain detailed data.

In their simplest formulation markovian mechanistic mobility processes are
equivalent to the classic reaction diffusion processes used in many
physical, chemical and biological processes~\citep{Marro:1999,Kampen:1981,murray}.  
The reaction-diffusion framework~\citep{Colizza:2007b} considers that 
the occupation numbers $N_i$ of each subpopulation can have any integer value,
including $N_i=0$, that is, void nodes with no individuals. The total
population of the metapopulation system is $N=\sum_i N_i$ and each
individual diffuse along the edges with a diffusion coefficient
$d_{ij}$ that depends on the node degree, subpopulation size and/or the
mobility matrix. A sketch of the metapopulation model which shows the 
different scales of the system is shown in Fig.~\ref{fig:meta-ill}. 
The  system
is  composed of a network substrate connecting subpopulations
over which individuals diffuses. Each
subpopulation is represented by a node $i$ of the network.
We consider that each node $i$ is connected to other $k_i$
nodes according to its degree resulting in a 
network with degree distribution $P(k)$ and
distribution moments $\langle k^\alpha\rangle=\sum_k k^\alpha P(k)$. 

In the case of large metapopulation systems with a
high level of heterogeneity the analytical description of the
metapopulation model in terms of specific features of each single 
subpopulation is extremely complicate. In the following we propose
an analytical framework that uses degree block variables to obtain the
dynamical equations describing the system's behavior, 
relying on the empirical evidence pointing to a statistical
equivalence of subpopulations having the same degree.

\begin{figurehere}
\begin{center}
\includegraphics[width=8.5cm]{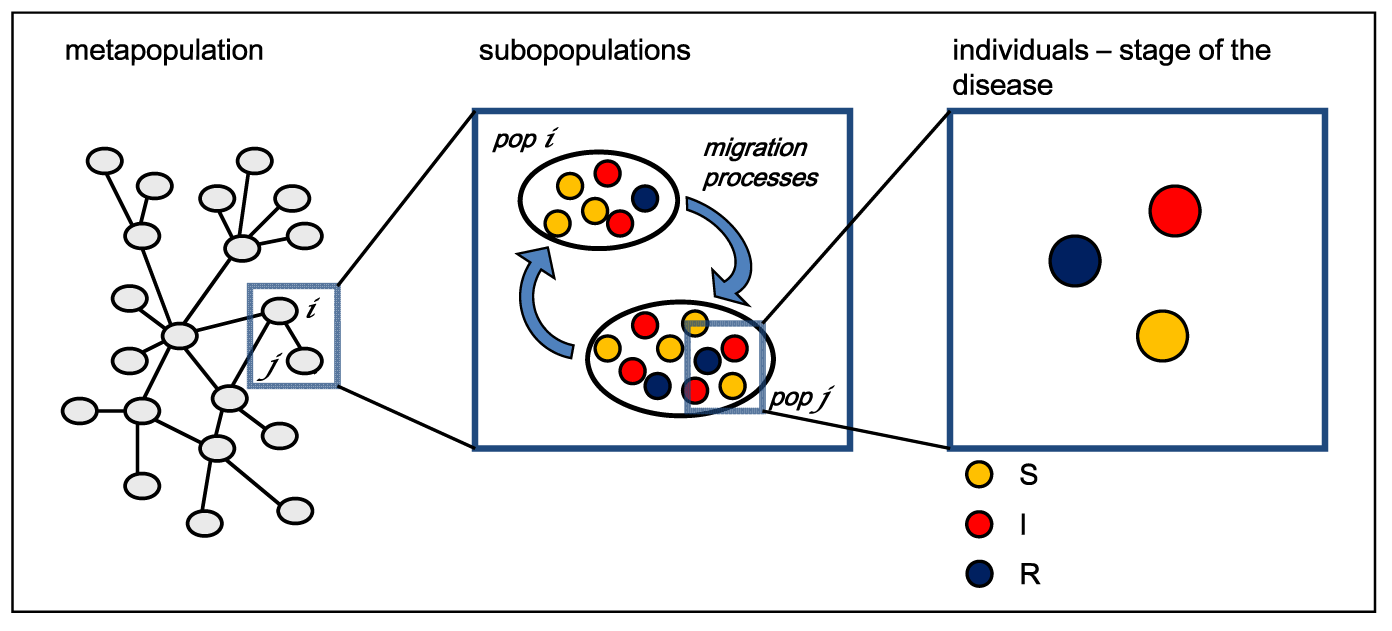}
\renewcommand{\baselinestretch}{1.0}
\caption{\small Schematic representation of a metapopulation model.
The system is composed of a heterogeneous network of subpopulations or patches,
connected by migration processes. Each patch contains a population of 
individuals who are characterized with respect to their stage of the disease 
(e.g. susceptible, infected, removed), and identified with a different color in the picture. 
Individuals can move from a subpopulation to another on the network of connections among subpopulations.}
\label{fig:meta-ill}
\end{center}
\end{figurehere}

%*****************************************************************************
\Subsection{Metapopulation networks with heterogeneous topology}
\label{heterogeneity}
In the real world, the network specifying the coupling between
different subpopulations is in many cases very heterogeneous. Examples
can be drawn from several transportation infrastructures, commuting
data and census information~\citep{Chowell:2003,transims,Barrat:2004,
Guimera:2005,DeMontis:2007}. 
A particularly relevant one in the field of
epidemic modeling is given by the airline transportation network. In
this case the coupling is provided by the number of passengers
traveling on a given route connecting two airports,  thus yielding a transfer of
 individuals between the corresponding urban areas. 
For instance, \cite{Barrat:2004} reports a detailed study of the International Air
Transport Association\footnote{IATA, International Air Transport Association, http://www.iata.org/} database which contains the 
complete list of world commercial airport  pairs connected by direct
flights. Moreover, to each direct flight connection between airports
$j$ and $\ell$ is assigned a weight $w_{j\ell}$ which corresponds to 
the number of available seats or passengers on the given route. 
The obtained network displays high levels of heterogeneity both in the
connectivity pattern and in the traffic capacities, as revealed by the
broad distributions of the number of connections of each airport, of
the travel flows between connected airports and of the traffic in
terms of number of passengers handled by each
airport~\citep{Barrat:2004}. These results have been confirmed on 
the US subnetwork along different
years and considering both market and segment traffic 
data\footnote{BTS, Bureau of Transportation Statistics, http://www.bts.org/}   
and analogous results are recovered by analyzing commuting patterns data, 
intra-city traffic among locations, and several other data
sets concerning the movements of people 
and goods~\citep{Chowell:2003,transims,Barrat:2004,Guimera:2005,DeMontis:2007}. In
Fig.~\ref{fig:weightednet} we report the degree  and weight  probability
distributions in some examples of these networks. In many cases we find
heavy-tailed distributions varying over  several orders of magnitude.
For instance the airline traffic among different urban areas in the world
shows a probability distribution $P(w)$~-~where $w$ is the traffic on a single
connection~-~varying over six orders of magnitude 
(see Fig.~\ref{fig:weightednet}B). 
\begin{figurehere}
\begin{center}
\includegraphics[width=8.5cm]{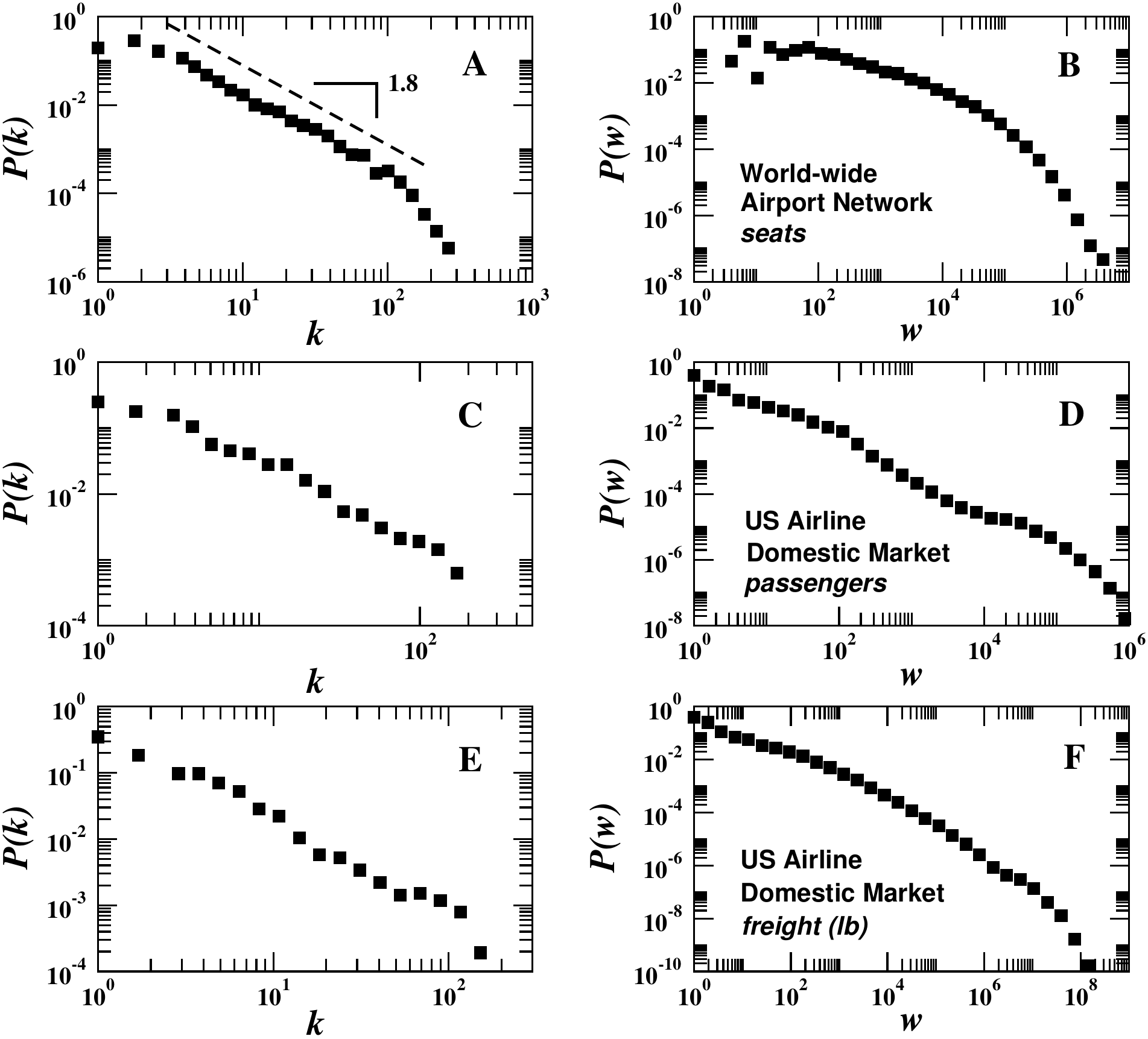}
\renewcommand{\baselinestretch}{1.0}
\caption{\small Degree (left column) and weight (right column) 
probability distributions for three
different datasets of the movement of people and goods: (A)-(B) world-wide
airport network, where the weight  represents the number of seats on the 
flights connecting two different airports (source: IATA); (C)-(D) US Air Domestic Market,
where the weight  represents the number of passengers
flying on a  given origin-destination itinerary (source: BTS); (E)-(F) US Air Domestic
Market for freight transportation, where the weight represents the amount
of freight (expressed in lb) transported from an origin airport to its 
final destination airport (source: BTS). All datasets show heavy-tailed distributions both in the number of connections and in the amount of people/good transported.}
\label{fig:weightednet}
\end{center}
\end{figurehere}

In addition, it is also possible to find some general statistical laws
relating the traffic and the degree of each node in the network. In
general the behavior of the average weight along the connection
between two subpopulations
with degree $k$ and $k'$ is
a function of their degree
\begin{equation}
  \langle w_{kk'}\rangle=w_0(kk')^{\theta}\,
\label{eq:w_kk'}
\end{equation}
where $w_0$ and the exponent $\theta$ depend on the specific
system (e.g. $\theta \simeq 0.5$ in the world-wide air transportation
network~\citep{Barrat:2004}). A related quantity is the total average traffic  
$T_k$ of the subpopulations with degree $k$ that behaves as 
\begin{equation}
  T_k= A k^{(1+\theta)}.
\end{equation}
Here the proportionality constant $A$ and the exponent $1+\theta$ are
defined by the sum rule $T_k=\sum_{k'}w_0(kk')^{\theta}$ that must
be satisfied on average. This last relation gauges the proportionality 
constant yielding $A= \langle
k^{1+\theta}\rangle w_0/\langle k \rangle$. It is important to stress
that the above relations defines a statistical equivalence of the
subpopulations of degree $k$. 
In the following we want to address the questions of how the large
scale complex features (scale invariance, extreme heterogeneity,
unbounded fluctuations) of interaction and communication networks
affect the behavior of metapopulation models by defining a
mechanistic approach based on the master equations describing the
disease dynamics as a microscopic reaction-diffusion process.
The equivalence assumption that will be used throghout the rest
of this paper is crucial in order to carry out the analytical treatment of the
model. While this assumption is indeed recovered in several data
sets, exceptions and fluctuations have been noted that would require more
complicate calculation schemes.

%*****************************************************************************
%*****************************************************************************
\Section{Mobility processes and diffusion properties in heterogeneous networks}
\label{diffusion}

In order to tackle the description of metapopulation models at the
mechanistic level, let us first analyze the simple diffusion process
of a global population of $N$ individuals who diffuse in a network
made of $V$ nodes, each representing a
subpopulation. Each node $i$ 
stores a number $N_i$ of individuals as defined in section 2.
In order to take into account the topological fluctuations of the
network we have to explicitly consider the presence of nodes with
a widely fluctuating degree $k$. A more convenient representation of the
system is therefore provided by the quantities  
\begin{equation}
N_k=\frac{1}{V_k}\sum_{i|k_i=k} N_i\,, 
\end{equation}
where $V_k$ is the number of nodes with degree $k$ and the sums run
over all nodes $i$ having degree $k_i$ equal to $k$.
The variable $N_k$ is therefore representing the average number of
individuals in subpopulations within the degree block $k$. This
representation is assuming that all subpopulations of the same degree
are statistically equivalent. 

Let us assume a general framework in which the individuals 
move from a subpopulation
with degree $k$ to another with degree $k'$ with a diffusion rate $d_{kk'}$
that depends on the degrees of the origin and destination subpopulations.
The probability of leaving a subpopulation with degree $k$
is then given by $p_k=\sum_{k'}P(k'|k)d_{kk'}$, where $P(k'|k)$ is the
conditional probability that any given edge departing from a node of degree $k$ is
pointing to a node of degree $k'$.
In the following, we will first write the equations for the 
dynamics of the individuals under this generic type of diffusion, and then
address two specific diffusion rates to find the stationary solutions.

The dynamics of individuals is simply represented by a mean-field
dynamical equation expressing the variation in time of the subpopulations 
$N_k(t)$ in each degree block. This can be easily written as: 
\begin{equation}
\pt N_k(t)= -p_k N_k(t) + k\sum_{k'}P(k'|k)d_{k'k}N_{k'}(t).
\label{diffeq}
\end{equation}
The first rhs term of the equation just considers that only a fraction
of particles $p_k$ moves out of the node. The second term instead
accounts for the particles diffusing from the neighbors into the
node of degree $k$. This term is proportional to the number of links
$k$ times the average number of particles coming from each
neighbors. This is equal to average over all possible degrees $k'$ the
fraction of particles moving on that edge 
$d_{k'k}N_{k'}(t)$ according to the  conditional probability
$P(k'|k)$. In the following we will consider
the case of uncorrelated networks in which the conditional
probability does not depend on the originating node, i.e.
 $P(k'|k)= k'P(k')/\langle k \rangle$~\citep{mendesbook,Pastor:2001}. 
This relation
simply states that any edge has a  probability to
point to a node with degree $k'$ that is proportional to the degree of
the node.
By using this form for $P(k'|k)$, 
the dynamical rate equation~(\ref{diffeq}) 
for the subpopulation densities reads as
\begin{equation}
\pt N_k(t)= -p_k N_k(t) + \frac{k}{\langle k\rangle}\sum_{k'}k'P(k')d_{k'k}N_{k'}(t).
\end{equation}
In the following subsections we solve the previous set of equations
for different diffusion processes that consider diffusion rates 
depending on the traffic of each node or on the population size of each
subpopulation.

%*****************************************************************************
\Subsection{Traffic dependent mobility rates}
Here
we assume that the probability an individual leaves 
a given subpopulation  is independent of its degree 
$k$, $p_k=p\,\,\,\forall k$. If we also assume homogeneous
 diffusion along any given connection, individuals have the same probability 
to move along anyone of the links
departing from the node at which they are located. In this case the diffusion
rate along any given link of a node with degree $k$ 
will be simply equal to 
\begin{equation}
d_{kk'}=p/k.
\end{equation}
This is obviously not the
case in a wide range of  real systems where 
the extreme heterogeneity of traffic is
well documented (see subsection 2.1). A more realistic
process therefore considers the movement of individuals to be
proportional to the traffic intensity along a given edge. This is simply
obtained by defining a heterogeneous diffusion probability for any
given individual to go from a subpopulation of degree $k$ to a
subpopulation of degree $k'$ as
\begin{equation}
d_{kk'} = p \frac{w_0(kk')^{\theta} }{T_{k}}.
\label{hetcoeff}
\end{equation}
This relation states that the diffusion rate $p$ is still constant in each
subpopulation but the individuals move on each connection in a proportion
dependent on the actual traffic on the connection. The denominator 
$T_{k}=Ak^{(1+\theta)}$ provides the correct normalization in order to
ensure that by summing over all $k$ edges 
departing from the node the overall diffusion rate is $p$. 

By using  the
expression of eq.~(\ref{hetcoeff}) for $d_{kk'}$ and imposing $p_k=p$, 
the dynamical rate equation~(\ref{diffeq}) 
for the subpopulation densities reads as
\begin{equation}
\pt N_k(t)= -p N_k(t) + p k^{(1+\theta)}\frac{w_0}{A \langle k
\rangle}\sum_{k'} P(k')N_{k'}(t).
\end{equation}
The stationary solution $\pt N_k(t)=0$ does not depend upon 
the diffusion rate $p$ that just fixes the time scale at which the
equilibrium is reached and has the solution 
\begin{equation}
N_k = k^{(1+\theta)}\frac{w_0}{A \langle k
\rangle}\bar{N},
\label{eq:Nk}
\end{equation}
where  $\bar{N}=\sum_{k}
P(k)N_{k}(t)$ represents the average subpopulation size. 
The explicit form of the normalization constant $A= \langle
k^{1+\theta}\rangle w_0/\langle k \rangle$, finally provides the explicit
stationary solution 
\begin{equation}
N_k = \frac{k^{(1+\theta)}}{\langle k^{(1+\theta)}
\rangle}\bar{N}.
\label{statdiff}
\end{equation}
The above solution states that the population 
of each node scales with the node degree in the stationary limit.  
The above behavior is simply the effect of the diffusion process that brings a
large number of individuals in well connected, high traffic nodes, 
thus showing the impact of network's topological (i.e. dependence on $k$) 
and traffic (i.e. dependence on $\theta$)
fluctuations  on the  individuals density behavior. 
When $\theta=0$ we recover the homogeneous diffusion case in which
$d_{kk'}=d_{k}=p/k$, obtaining
\begin{equation}
N_k = \frac{k}{\langle k \rangle} \bar{N}.
\end{equation}
In this case the subpopulation density is just fixed from topological
fluctuations and the exponent $\theta$ clearly appears as the
parameter that takes into account the traffic fluctuations.
It is worth remarking that in this framework, 
the subpopulation size as a
function of the degree is constrained by
the diffusion processes, a feature that has not to be expected in real
systems where the population size of local patches
can be considered as an independent variable. 
On the other hand, the degree dependence is close to those observed in
real systems where in several cases it is possible to find a relation
$N_k\sim k^\phi$ with $0.5\leq\phi\leq 1.5$~\citep{Colizza:2006a,Colizza:2006b}. We can thus
consider the obtained stationary state as a first approximation to the
real case and use the exponent $\theta$ to explore different levels of
heterogeneity. 

%*****************************************************************************
\Subsection{Population dependent mobility rates}
\label{subsec:diff_pop_dep}
In a more general perspective, it is important to have the possibility of
considering the population densities $N_k$ as independent
variables. This is indeed the case of many metapopulation models in which
the diffusion process represents the travel of
individuals between subpopulations. In this framework the number of
people traveling from a subpopulation to the other in a unitary time
scale  is a defined number
$w_{ij}$ and the number of traveling individuals is independent from
the population size $N_i$. This  amounts to state that each individual
in the subpopulation has a diffusion rate $\sum_j w_{ij}/N_i$ where
$\sum_j w_{ij}$ is the total number of
people traveling out of city $i$ in the unitary time scale. In other
words, the diffusion rate of each individual is inversely proportional
to the population size. In order to have a non-pathological stationary
state the condition $w_{ij}=w_{ji}$ has to be satisfied at least on
average. In this case we can write
\begin{equation}
\pt N_i = \sum_j \left(w_{ji}-w_{ij}\right)=0,
\label{eq:const_pop}
\end{equation}
and any initial conditions for the population size satisfies the
stationary state. 
In the degree block variable representation we can recover the above
condition by considering a diffusion rate for each particle of the form
$p_k=T_k/N_k$. The diffusion rate on any given edge from a subpopulation
of degree $k$ to a subpopulation of degree $k'$ is therefore given by 
\begin{equation}
d_{kk'} = \frac{w_0(kk')^{\theta}}{N_{k}}
\label{eq:pop_coeff}
\end{equation}
and the degree block diffusion equations read in the uncorrelated
networks case as
\begin{equation}
\pt N_k(t)= -T_k + k^{(1+\theta)}w_0\frac{\langle k^{1+\theta}\rangle}{\langle k \rangle}.
\end{equation}
Since we know that by normalization $T_k=
k^{(1+\theta)}w_0\langle k^{1+\theta}\rangle/\langle k\rangle$, we recover by
definition the solution $\pt N_k(t)=0$ that allows any stationary
value distribution $N_k$. Differently from the results obtained in the
previous subsection, where 
each individual has the same probability $p$ of leaving a subpopulation,
eq.~(\ref{eq:const_pop}) shows that a population dependent diffusion process does not 
fix the subpopulation size, which can be given as a parameter of the
model, with the only constraint that $N_k>T_k$, in order to make the diffusion 
process feasible.

%*****************************************************************************
%*****************************************************************************
\Section{Epidemic spreading and the invasion threshold}
\label{epidemic}

In order to explore the epidemic behavior in metapopulation models, 
the disease dynamics needs to be explicitly considered
inside each subpopulation. In the
following we will consider the standard compartmentalization approach in 
which individuals 
exist in a certain number of  discrete states such as  susceptible, 
infected or permanently recovered~\citep{anderson92}. 
The paradigmatic epidemiological model one can consider, is the
susceptible-infected-removed (SIR) model~\citep{anderson92,murray}, where the
total number of individuals 
$N_j$ in the subpopulation $j$ is partitioned in the compartment 
$S_j(t)$, $I_j(t)$, and $R_j(t)$  denoting 
the number of susceptible, infected and recovered individuals at
time $t$, respectively. By definition it follows $N_j=S_j(t)+I_j(t)+R_j(t)$. 
The disease transmission is described in an effective way.  The
probability that a susceptible individual acquires the infection from any
given neighbor in an infinitesimal time interval $d t$ is
$\beta d t$, where $\beta$ defines the 
disease \index{spreading rate}\textit{transmissibility}.  At the 
same time, infected vertices are cured and become
recovered with probability $\mu d t$.
Individuals thus run stochastically through the susceptible $\to$
infected $\to$ recovered transitions, hence the name of the model.
The SIR model assumes that recovered individuals are basically removed
from the system, they do not participate anymore to the disease
dynamics, due to their death or acquired immunization. Another popular
model takes into account the possibility that infected individuals 
are again susceptible with probability $\mu d t$. In this case 
individuals thus run stochastically through the cycle susceptible $\to$
infected $\to$ susceptible, defining the so-called SIS model.
The SIS model is mainly used as a
paradigmatic model for the study of infectious diseases leading to an
endemic state with a stationary and constant value for the prevalence
of infected individuals, i.e. the degree to which the infection is
widespread in the population. 

A basic parameter in the analysis of a single population epidemic
outbreaks is the basic reproductive 
number $R_0$, which counts the number of secondary infected cases 
generated by a primary infected individual~\citep{anderson92}.
Under the assumption of the homogeneous mixing of the population 
the basic reproductive number is defined as
\begin{equation}
  R_0 = \frac{\beta}{\mu}.
\label{eq:R0}
\end{equation}
It is straightforward to see from eq.~(\ref{eq:R0}) that in the single
population case any epidemic will
spread across a non zero fraction of the population only for $R_0>1$.
In this case the epidemic is able to generate a number of infected
individuals larger than those who recover,  leading to an
increase in the overall number of  infectious individuals $I(t)$. 
The previous considerations lead to the definition of a crucial
epidemiological concept~-~the epidemic threshold~\citep{anderson92}. Indeed, if the
spreading rate is not large enough  to allow a reproductive number
larger than one (i.e. $\beta>\mu$), the epidemic outbreak will not
affect a finite portion of the population and will die out in a finite
amount of time.  

At the metapopulation level, however, the epidemic behavior on the
global scale is determined also by the diffusion process of
individuals. In particular,  the effects due to the finite size of 
subpopulations and the stochastic nature
of the diffusion might have a crucial
role in the problem of resurgent epidemics, extinction and 
eradication~\citep{Ball:1997,Cross:2005,Watts:2005,Vazquez:2007,Cross:2007}. 
Therefore it is
important to consider the discrete nature of individuals.  Indeed, each
subpopulation may or may not transmit the infection to another subpopulation it
is in contact with, 
depending on the occurrence or not of the travel event of at least one
infected individual  to the non-infected subpopulation during the
entire epidemic evolution. 
The spreading process across subpopulations will therefore 
occur with a probability that is related to the
diffusion probability of individuals and the total number of
individuals that will experience the 
infection~\citep{Ball:1997,Cross:2005,Cross:2007}. In the case of
epidemic processes with $R_0<1$ the epidemic will die with probability
1 and  is not going to spread across subpopulations. 
In a model like the
SIS model, if $R_0>1$ the number of infected individuals reaches a
stationary state and the epidemic will eventually spread to different
subpopulations, since locally endemic. 
In a model such as the SIR model, however, the epidemic
within each subpopulation generates a finite fraction of infectious
individuals in a given amount of time and even if $R_0>1$ 
the diffusion rate must be large enough to ensure the timely diffusion
of infected individuals to other subpopulations of the metapopulation
system, before the local epidemic outbreak dies out. This is captured
by the definition of a new predictor of disease invasion, $R_*$, regulating
the number of subpopulations that become infected from a single
initially infected subpopulation; i.e. the analogous of the
reproductive number at the subpopulation level~\citep{Ball:1997,Cross:2005,Cross:2007}.

This effect would not be captured by a simple deterministic
description that would allow any fraction $pI$ of diffusing infected 
individual to inoculate the virus in a subpopulation not yet infected.
In certain conditions this fraction $pI$ may be a number
smaller than one that just represents a mean-field average value. This
is a common error of deterministic continuous approximations that
allow the infection to persist and diffuse via ``nano-individuals''
that are not capturing the discrete nature of the real systems.
For this reason, in the next section we will use an approach working
at the level of subpopulations that allows to take into account effectively the
fluctuations inherent to the diffusion process and the outbreak
extinction probability.

%*****************************************************************************
\Subsection{Global invasion threshold in homogeneous metapopulation networks}

Let us consider a metapopulation system in which the initial 
condition is provided by a single introduction in a subpopulation of 
degree $k$ and size $N_k$, given  $R_0>1$. While the stochastic
nature of the process may lead in some cases to the extinction of the
process, as $R_0$ is above the epidemic threshold the epidemic will
affect a finite fraction of the population with non zero probability.
In the case of a macroscopic outbreak in a closed population the total number
of infected individuals during the evolution of the epidemic will be
equal to $\alpha N_{k}$ where  $\alpha$ depends on the specific
disease model used and the disease parameter values. Each infected individual
stays in the infectious state for an average time $\mu^{-1}$ equal to the
inverse of the recovery rate, 
 during which it can travel to the neighboring subpopulation of degree $k'$ 
with rate $d_{kk'}$.
To a first approximation we can therefore consider that the
number of new seeds that may appear into a connected
subpopulation of degree $k'$ during the 
duration of the subpopulation epidemic is given by 
\begin{equation}
\lambda_{kk'}=d_{kk'}\frac{\alpha N_k}{\mu}.
\end{equation}
In this perspective we can consider the metapopulation model in a
coarse grained view (see Fig.~\ref{fig:invasion}) and provide a 
characterization of 
the invasion dynamics at the level of the subpopulations, translating
epidemiological and demographic parameters into Levins-type metapopulation
parameters of extinction and invasion rate.
Let us define $D^0_k$ as the number of \emph{diseased} 
subpopulation of degree $k$ at generation $0$, i.e. those which
are experiencing an outbreak at the beginning of process. Each
infected subpopulation during the course of the outbreak will seed the 
infection in neighboring subpopulations defining the set $D^1_k$ of infected
subpopulations at the following generation  and so on.
This corresponds to a basic branching
process~\citep{Harris:1989,Ball:1997,Vazquez:2006} where the $n-$th 
generation of infected subpopulations of degree $k$ 
is denoted $D^n_k$.  
\begin{figurehere}
\begin{center}
\includegraphics[width=8.2cm]{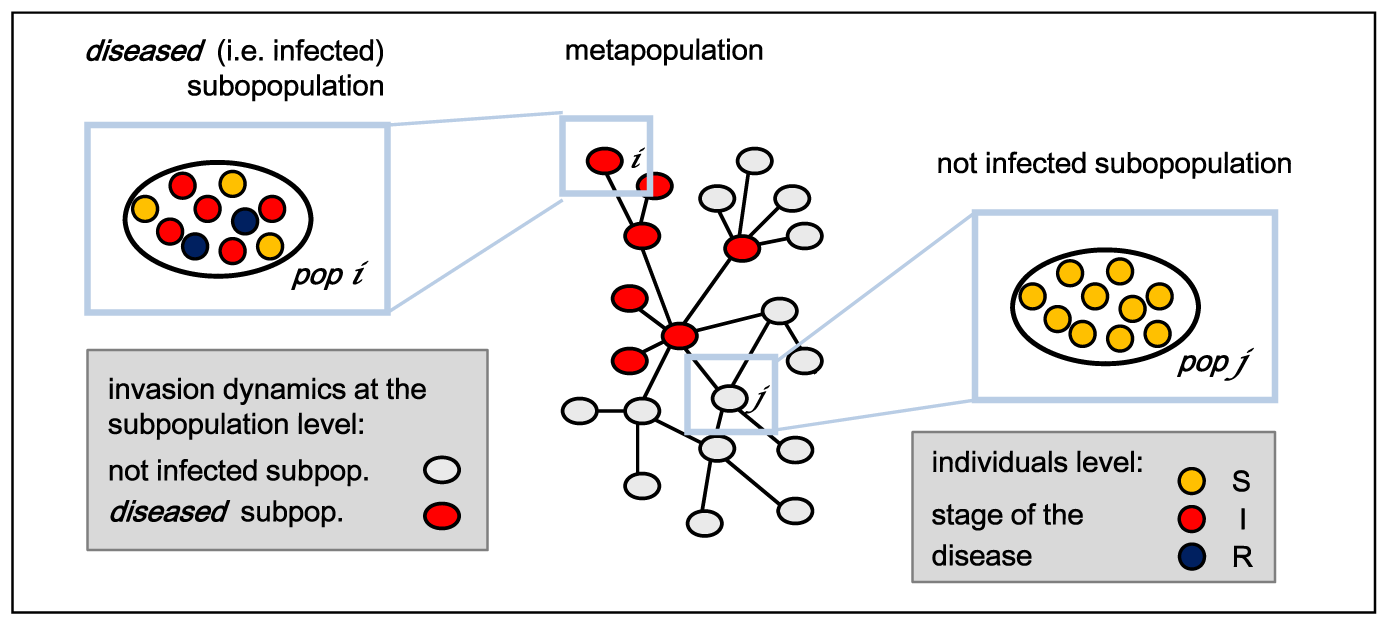}
\renewcommand{\baselinestretch}{1.0}
\caption{\small Schematic representation of the invasion dynamics at the level of 
the subpopulations. The metapopulation system can be considered in a coarse grained
perspective as a network where each node represents a subpopulation which can be 
infected (i.e.
\emph{diseased}) if it is reached by the virus as carried by the infected 
individuals diffusing on the system.}
\label{fig:invasion}
\end{center}
\end{figurehere}

In the early stage of the epidemics we assume that the number of
subpopulations affected by an outbreak (with $R_0>1$) is small and we can 
therefore study the
evolution of the number of diseased subpopulations by using a tree-like
approximation relating $D^n_k$ with $D^{n-1}_k$. Let us
first analyze the case of a metapopulation system in the form of a
homogeneous random graph in which each subpopulation has the same degree
$k=\bar{k}$ and population $\bar{N}$. In this case we can drop the
subscript index $k$ (all subpopulations being equal) and  obtain that 
\begin{equation} 
D^n = D^{n-1} (\bar{k}-1)
\left[1-\left( \frac{1}{R_0}\right)^{\lambda_{\bar{k}\bar{k}}}\right]
\left( 1-\frac{D^{n-1}}{V} \right).
\label{poptree}
\end{equation}
This equation assumes that each infected subpopulation 
of the $(n-1)-$th generation,
$D^{n-1}$, will seed with infected individuals a number of subpopulations 
during the course of the outbreak
 that
depends on the product of: the number of neighbor subpopulations minus the one 
which originally transmitted the disease, $\bar{k}-1$, times 
the probability that the subpopulation is not already seeded
by infected individuals,  $(1-D^{n-1})/V$, and the probability that
the new seeded subpopulation will experience an outbreak, 
i.e. $\left(1-R_0^{-\lambda_{\bar{k}\bar{k}}}\right)$~\citep{Bailey:1975}. The last
expression stems from the probability of extinction $P_{ext}=1/R_0$ 
given by a seed of a single infectious individual~\citep{Bailey:1975}.
The simplest case of homogeneous
diffusion of individuals $d_{\bar{k}}=p/\bar{k}$ yields 
$\lambda_{\bar{k}\bar{k}}= p \bar{N}\alpha\mu^{-1}/\bar{k}$. In order to obtain an
explicit result we will consider in the following that $R_0-1\ll 1$,
so that the system is poised very close to the epidemic threshold. 
In this limit we can approximate the outbreak probability as
\begin{equation}
\left[1-\left(\frac{1}{R_0}\right)^{\lambda_{\bar{k}\bar{k}}}\right]
\simeq \lambda_{\bar{k}\bar{k}}(R_0-1),
\end{equation}
 and assuming that at the early stage of the epidemic
$D^{n-1}/V \ll 1$ we obtain
\begin{equation} 
D^n =p\bar{N}\alpha\mu^{-1}\frac{\bar{k}-1}{\bar{k}}(R_0-1)D^{n-1}. 
\end{equation}
This relation states that the number of subpopulations affected by an
outbreak will increase only if the quantity 
\begin{equation} 
R_*=p\bar{N}\alpha\mu^{-1}\frac{\bar{k}-1}{\bar{k}}(R_0-1)>1.
\end{equation}
This relation defines the \emph{global invasion threshold}, i.e. the 
subpopulation reproductive number $R_*$ that
is the analogous of the basic reproductive number $R_0$ in structured
metapopulation models. From the above expression it is
possible to write the threshold condition on the mobility rate 
\begin{equation} 
p\bar{N}\geq \frac{\bar{k}}{\bar{k}-1}\frac{\mu}{\alpha}(R_0-1)^{-1}, 
\end{equation}
that fixes the threshold in the diffusion of individuals 
for the global spread of the epidemic in the metapopulation systems. 
In other words, this equation states that there is a minimum rate 
for the traveling
of individuals in order to ensure that on average each subpopulation can
seed more than one neighboring subpopulations. As it has been pointed
out by~\cite{Cross:2007} we find that other factors such as the
infectious period and the mobility process are as much important as
$R_0$ in the spread of epidemics in structured populations. 
The constant $\alpha$ is larger than zero for any
$R_0>1$, and in the SIR case for $R_0$ close to 1 it can be approximated 
by~\citep{murray}:
\begin{equation} 
\alpha\simeq
2\frac{\mu}{\beta}(1-\frac{\mu}{\beta})=\frac{2(R_0-1)}{R_0^2},
\end{equation}  
yielding the mobility  threshold for the SIR model 
\begin{equation} 
p\bar{N}\geq \frac{\bar{k}}{\bar{k}-1}\frac{\mu R_0^2}{2(R_0-1)^2}.
\label{glob}
\end{equation}
The above condition readily tells us that the closer to the epidemic
threshold is the epidemic in the single subpopulation and the larger it has
to be the traveling rate  in order to sustain the global spread into 
the metapopulation model. We therefore find that we can define two
different thresholds in a homogeneous metapopulation model. The first
one is the local epidemic threshold $R_0>1$ within each subpopulation and
the second one $R_*>1$ represents the global invasion threshold defining the
traveling rate of individuals according to eq.~(\ref{glob}). It is
important to stress that when $R_0$ increases the small $R_0-1$
expansions are no longer valid and the invasion threshold is obtained
only in the form of a complicate implicit expression.

%*****************************************************************************
\Subsection{Global invasion threshold in metapopulation networks
with traffic dependent mobility rates}
\label{subsec:glth_traffic_diff}
The calculation for the global threshold becomes more complicate
in the case of heterogeneous metapopulation networks. 
In this case eq.~(\ref{poptree}) has to include also the degree and
population heterogeneities yielding:
\begin{equation} 
D_k^n = \sum_{k'}D_{k'}^{n-1} (k'-1) 
\lambda_{k'k}(R_0-1) P(k|k')\left(1-\frac{D_k^{n-1}}{V_k}\right).
\label{poptree-het}
\end{equation}
This expression considers that each subpopulation of degree $k'$ will seed the
infection in a number $k'-1$ of subpopulations corresponding  to 
the number of neighboring subpopulations minus the one which originally
transmitted the infection, 
the probability $P(k|k')$ that each of the $k'$ neighboring 
populations has degree $k$, and the probability to observe an
outbreak in the seeded population, where as before we considered the
limit $R_0-1\ll 1$ to obtain the the outbreak probability 
as $\lambda_{k'k}(R_0-1)$. As in the previous case of the homogeneous 
network, we consider the
early stage of the epidemic in which $(1-D_k^{n-1}/V_k)
\simeq 1$. In addition we assume that degree correlations can be neglected and 
$P(k|k')= kP(k)/\langle k \rangle$ obtaining
\begin{equation} 
D_k^n =\frac{kP(k)}{\langle k \rangle}(R_0-1)\sum_{k'}D_{k'}^{n-1} (k'-1) \lambda_{k'k}.
\end{equation}
The behavior of the above expression depends on the specific form of
$\lambda_{k'k}$ that is determined by the diffusion rate $d_{kk'}$. Let us
first consider the heterogeneous diffusion rate of
Eq.~(\ref{hetcoeff}) that gives
\begin{equation}
\lambda_{k'k}=\frac{p\langle k \rangle}{\langle k^{(1+\theta)}\rangle}
\frac{\alpha}{\mu} \frac{(k)^\theta}{k'}N_{k'}=
\frac{p\langle k \rangle}{\langle k^{(1+\theta)}\rangle^2}
\frac{\alpha}{\mu}(kk')^\theta\bar{N},
\end{equation}
where in the last expression we have considered that the size of each
population of degree $k$ is given by the stationary diffusion process
according to Eq.~(\ref{statdiff}). The equation describing the
generation of infected subpopulations is therefore reading as
\begin{equation} 
D_k^n =(R_0-1)\frac{k^{1+\theta}P(k)}{\langle k^{1+\theta} \rangle^2}
\frac{p\bar{N}\alpha}{\mu} \sum_{k'}D_{k'}^{n-1} k'^{\theta}(k'-1).
\end{equation}
By defining $\Theta^n= \sum_{k'}D_{k'}^{n} k'^{\theta}(k'-1)$ the 
last expression can be conveniently written in the iterative form
\begin{equation} 
\Theta^n =(R_0-1)\frac{\langle k^{2+2\theta} \rangle-\langle k^{1+2\theta} \rangle}{\langle k^{1+\theta} \rangle^2}
\frac{p\bar{N}\alpha}{\mu} \Theta^{n-1},
\end{equation}
that allows the increasing of infected subpopulations and a global
epidemic in the metapopulation process only if 
\begin{equation} 
R_*=(R_0-1)\frac{\langle k^{2+2\theta} \rangle-\langle k^{1+2\theta} \rangle}{\langle k^{1+\theta} \rangle^2}
\frac{p\bar{N}\alpha}{\mu} >1.
\end{equation}
The subpopulation reproductive number is therefore an increasing function of the
network heterogeneity that plays a role in the spread of the pathogen 
across subpopulations.
In the case of an SIR epidemic within each subpopulation, the
threshold on the mobility rate is provided by the expression
\begin{equation} 
p\bar{N}\geq \frac{\langle k^{1+\theta} \rangle^2}{
\langle  k^{2+2\theta} \rangle-\langle  k^{1+2\theta} \rangle}\frac{\mu R_0^2}{2(R_0-1)^2},
\label{eq:glth1}
\end{equation}
that differs from the homogeneous case for a correction factor
depending on the topology of the network. Noticeably, the ratio
$\langle k^{1+\theta} \rangle^2/(\langle k^{2+2\theta}\rangle)-\langle k^{1+2\theta}\rangle)$ is extremely
small in heavy-tailed networks and it is vanishing in the limit of
infinite network size. This implies that the heterogeneity of the
metapopulation network is favoring the global spread of epidemics by
lowering the global invasion threshold.

%*****************************************************************************
\Subsection{Global invasion threshold in metapopulation networks
with population dependent mobility rates}
As a final case let us consider the realistic framework in which the
diffusion rate of individuals is proportional to the ratio between
traveling people and population size; i.e. $p_k=T_k/N_k$. In 
subsection 3.2 we have seen that this case corresponds to the
mean-field assumption in metapopulation models coupled by traveling
fluxes, leading to a stationary state in which the population $N_k$ is
stationary and independent on the diffusion process. 
Here $\lambda_{kk'}= w_0(kk')^\theta\alpha\mu^{-1}$ and by using the
approximations considered in the previous cases
the basic equation for the number of infected subpopulations reads as
\begin{equation} 
D_k^n =(R_0-1)\frac{k^{1+\theta}P(k)}{\langle k \rangle}
\frac{w_0\alpha}{\mu} \sum_{k'}D_{k'}^{n-1} k'^{\theta}(k'-1).
\end{equation}
Also in this case by using the auxiliary function 
$\Theta^n= \sum_{k'}D_{k'}^{n} k'^{\theta}(k'-1)$ we obtain the
recursive relation 
\begin{equation} 
\Theta^n =(R_0-1)\frac{\langle k^{2+2\theta}\rangle-\langle k^{1+2\theta}\rangle}{\langle k \rangle}
\frac{w_0\alpha}{\mu} \Theta^{n-1},
\end{equation}
yielding for the global invasion the condition
\begin{equation} 
R_*=(R_0-1)\frac{\langle k^{2+2\theta}\rangle-\langle k^{1+2\theta}\rangle}{\langle k \rangle}
\frac{w_0\alpha}{\mu}>1. 
\end{equation}
In the case of an SIR model for the intra-population disease we obtain
\begin{equation} 
w_0\geq\frac{\langle k \rangle}{\langle k^{2+2\theta}\rangle-\langle k^{1+2\theta}\rangle}\frac{\mu
  R_0^2}{2(R_0-1)^2},
\label{eq:glth_pop}
\end{equation}
where also in this case the mobility threshold is lowered by the topological
fluctuations of the network as the more heterogeneous is the
metapopulation network and the smaller is the ratio 
$\langle k \rangle/(\langle k^{2+2\theta}\rangle-\langle k^{1+2\theta}\rangle)$. 
In this respect it is worth remarking that in principle 
in an infinite network with heavy-tails 
the mobility threshold is vanishing as the 
ratio $\langle k \rangle/(\langle k^{2+2\theta}\rangle-\langle k^{1+2\theta}\rangle) \to 0$. 
In an infinite network, however, the above equations should be rewritten 
in terms of the density of diseased subpopulations in order to avoid 
the pathological divergence of some terms. 
Finally, it is interesting to notice that the effect of the network
heterogeneity on the subpopulation reproductive number is similar to
that of the contact pattern heterogeneity on the basic reproductive
number~\citep{anderson92,pv01a,lloyd01,Barthelemy:2005}, 
stressing even more the close analogy between the two metrics.

%*****************************************************************************
\Subsection{Local and global threshold in real-world cases}
It is worth stressing that the previous expressions are approximate and
valid only in the limit in which  a small fraction of the
populations in the system is affected and in which $R_0-1\ll 1$. 
It is however extremely relevant that metapopulations systems have 
intrinsically two epidemic thresholds. 
The emergence of a global epidemic is first constrained
by the intrinsic epidemic threshold within each subpopulation,
$R_0>1$. If the epidemic process satisfies this condition, each time
an infectious individual seeds an epidemic within a subpopulation there is a
finite probability that a macroscopic fraction of the population will
be affected by the outbreak. While this condition guarantees the
intra-population spreading of the epidemic, the inter-population
spreading is controlled by the coupling among subpopulations as
quantified by the rate of diffusing/traveling individuals. The global
invasion threshold condition $R_*>1$ provides an estimate of the rate of
diffusion of individuals above which the epidemic is able to affect a
macroscopic fraction of the subpopulations defining the meta-population
network.

At this point, it is useful to draw some gross estimate of the critical
population coupling $w_0$ as a function of $R_0$ and of a realistic value
of $\mu$. If we assume a very mild reproductive rate of about
$R_0\simeq 1.1$ and a value $\mu=1/3$ per unit time (1 day) as in many estimates for
influenza strains, we obtain that $w_0$ per unit time must be larger
than approximately 20 individuals per day in
the case of homogeneous networks and one order of magnitude or more
smaller in heterogeneous networks. This is a value met in most of the
modern real transportation systems. For example, the world-wide air
transportation network analyzed in ref.~\citep{Barrat:2004}
and briefly described in subsection 2.1 is characterized by a topology
whose degree distribution moments which appear in the expressions of  the global
threshold are given by: $\langle k\rangle \simeq 10$, 
and 
$\langle k^{2+2\theta}\rangle-\langle k^{1+2\theta}\rangle \simeq 7\cdot 10^4$, given that
$\theta \simeq 0.5$~\citep{Barrat:2004}. Therefore the condition 
expressed in eq.~(\ref{eq:glth_pop}) states that
an epidemic carried by air travelers would reach global proportion
 if the average number of 
travelers per day is larger than approximately $3\cdot 10^{-3}$,
a constraint which is met in the airport network where the average daily
traffic on a given connection has a minimum corresponding to $\simeq
10^{-2}$\footnote{IATA, International Air Transport Association,
  http://www.iata.org/}.
The result of this estimate is in agreement with recent studies on contingency
planning for a possible influenza pandemic, which show that travel
restrictions, reducing the probability of any individual to leave an
infected region, would not be able to considerably slow down the
global spread unless $>90\%$ or more
effective~\citep{Hollingsworth:2006,Cooper:2006,Colizza:2007a}. Our
analytical results shows that the invasion threshold is extremely
small in realistic situations and traffic reduction of more than one
order of magnitude are in order to bring the system below the
threshold.

%*****************************************************************************
%*****************************************************************************

%*****************************************************************************
\Section{Epidemic behavior above the invasion threshold}

Above the invasion threshold we can assume that with finite
probability the epidemics will affect a macroscopic fraction of
subpopulations. In this limit, the stochastic effect due to the
diffusion can be neglected and it is possible to study the epidemic
spreading in the system by the deterministic equations obtained from a
mechanistic approach to the metapopulation model where the disease
dynamics in each subpopulation can be viewed as a 
reaction process~\citep{Colizza:2007b}. In
the case of the SIR scheme the dynamics is 
identified by the following set of reaction equations:
\begin{eqnarray}
  I + S &\to& 2 I \label{reaction1} \\
  I &\to& R.  \label{reaction2}
\end{eqnarray}
In the SIS case the second reaction is just replaced by the reaction
$I\to S$. From the rate equations it is clear that the
dynamics conserves the total number of individuals. 
Before the diffusion process, the $I_j$ and $S_j$ 
individuals belonging to the same subpopulation $j$ react according to 
the eqs.~(\ref{reaction1}) and (\ref{reaction2}). 
In each node $j$ the spontaneous process $I\to R$ 
simply consists in turning each $I_j$
individual into an $R_j$ individual with  rate $\mu$.
This process account for the recovery  of infected
individuals from the disease. 
The process $ I + S \to 2 I$ is related to
the dependence of the transmissibility on the population
density. In general, in large populations it is customary to consider that
each  individual has a finite number of contacts per unit
time. In this case the probability that 
a susceptible has a contact with an infectious individual 
 is equal to the density of infectious
individuals within the subpopulation $j$, i.e. $I_j/N_j$. If we consider 
a homogeneous mixing assumption within the
population,   the creation
rate of infectious individuals
will be provided by $\beta \Gamma_j$ where $\Gamma_j$ is an
interaction kernel of the form 
\begin{equation}
  \Gamma_j = \frac{I_j S_j}{N_j}.
\end{equation}
It is natural also to consider different dependencies of the
transmissibility with respect to the density, giving rise to different reaction
kernels~\citep{anderson92,Colizza:2007b}. We will provide an analysis  of the case of reaction kernels
simulating population with internal network structure and fully
connected populations in a forthcoming paper~\citep{PRE}.

%*****************************************************************************
\Subsection{Deterministic reaction-diffusion rate equations}
In order to provide the explicit equations describing the dynamical
evolution of the metapopulation system, we generalize the basic
mechanistic approach with degree block variables used 
in the previous section to the complete
reaction diffusion process. We take into account the topological 
fluctuations of the coupling 
networks by introducing the quantities:
\begin{equation}
I_k=\frac{1}{V_k}\sum I_j; ~~~~~~~~~~ S_k=\frac{1}{V_k}\sum S_j,
\end{equation}
which represent the the average number of $I$ and $S$ individuals
in subpopulations with degree $k$. Analogously, the reaction kernel in
the homogeneous assumption is written for subpopulation in each degree
block as $\Gamma_k=I_k S_k/N_k$. Again it is worth remarking that the
degree block variables assumes the statistical equivalence of
subpopulations with the same degree $k$. While this approximation is
in fair agreement with empirical analysis, real-world subpopulations
have differences that the present analysis does not take into account.

At the end of the reaction-diffusion 
process the variation in the number of infected individuals in subpopulations of
degree block $k$ can be written as a
discrete master equations with the form
\begin{equation}
 I_k(t+\Delta t)-I_k(t)= W^+_k - W^-_k,
\label{meq} 
\end{equation}
where the terms $W^+_k$, $W^-_k$ identify the number of infected
individuals that entered or left the class $I_k$ because of both the
disease dynamics and the diffusion process. Assuming the general
framework in which the diffusion probability out of a given
subpopulation depends on its degree, $p_k$,  the depletion
term $W^-_k$ can be simply evaluated as:
\begin{equation}
 W^-_k=p_k I_k + (1-p_k) \mu I_k.
\end{equation}
The depletion term is just the sum of the $I_k$ individuals that
diffuse away of the subpopulation (first term of the r.h.s.) 
and the infected individuals that do
not diffuse away from the subpopulation but have a transition to the class 
$R_k$ (second term of the r.h.s.). The positive
term $W^+_k$ takes into account both the new infected individuals
generated by the disease dynamics within the subpopulation and the
infected individuals that diffuse from the neighboring subpopulations
with diffusion rate $d_{kk'}$,
and it is given by:
\begin{equation}
 W^+_k= (1-p_k)\beta\Gamma_k + k\sum_{k'}P(k'|k)d_{k'k}\left[
(1-\mu)I_{k'}+\beta\Gamma_{k'}\right].
\end{equation}
The first term on the r.h.s. considers the newly generated infected
individuals within the subpopulation with degree $k$. The factor $1-p_k$
takes into account only those individuals that do not diffuse out of
the subpopulation. The second term accounts for all the infected
individuals arriving because of the diffusion process from neighboring
subpopulations. The factor $k$ considers the presence of $k$ neighboring
subpopulations, each one contributing a
fraction $d_{k'k}$ of its number of diffusing infected individuals 
which is given by the non recovering plus the newly generated ones
$(1-\mu)I_{k'}+\beta\Gamma_{k'}$. Finally, on each connection edge we have 
to average over the probability that the
neighboring subpopulation has degree $k'$, that is given by the weighted
sum over the conditional probability $P(k'|k)$.
As for the simple diffusion in section 3, we consider an infinitesimal
time interval $\Delta t \to 0$ and divide both terms of
equation~(\ref{meq}) to obtain the following set of differential equations 
\begin{eqnarray}
\pt I_k= -p_k I_k + (1-p_k)\left[-\mu I_k+\beta\Gamma_k\right]+~~~~\nonumber\\
~~~~+k \sum_{k'}P(k'|k)d_{k'k}\left[
(1-\mu)I_{k'}+\beta\Gamma_{k'}\right],
\end{eqnarray}
where all the parameter combinations are now infinitesimal transition
rates. 
By considering the uncorrelated case $P(k'|k)= k'P(k')/\langle
k \rangle$, we obtain the following dynamical reaction-rate equations
\begin{eqnarray}
\pt I_k= -p_k I_k + (1-p_k)\left[-\mu I_k+\beta\Gamma_k\right]+~~~~~~\nonumber\\
~~~~~~+\frac{k}{\langle k\rangle} \sum_{k'}k'P(k')d_{k'k}\left[
(1-\mu)I_{k'}+\beta\Gamma_{k'}\right].
\label{eq:unif_diff}
\end{eqnarray}
Similar expressions can be written for the evolution of $S_k$ and $R_k$
as:
\begin{eqnarray}
\pt S_k= -p_k S_k - (1-p_k)\beta\Gamma_k+~~~~~~~~~\nonumber\\
~~~~~~~~~+\frac{k}{\langle k\rangle} \sum_{k'}k'P(k')d_{k'k}\left[
S_{k'}-\beta\Gamma_{k'}\right],
\end{eqnarray}
and
\begin{eqnarray}
\pt R_k= -p_k R_k + (1-p_k)\mu I_k +~~~~~~~~~\nonumber\\
~~~~~~~~~+\frac{k}{\langle k\rangle} \sum_{k'}k'P(k')d_{k'k}\left[
\mu I_{k'}+R_{k'}\right].
\end{eqnarray}

%*****************************************************************************
\Subsection{The early stage of the  epidemic outbreak}
An explicit solution to the previous equations can be obtained for the
early stages of the epidemic. In this case we can imagine to have a
very small densities of infectious individuals in the metapopulation
system so that in general we can neglect contributions of order
$I_k^2$. In this setting the reaction kernel $\Gamma_k$ can be
approximated as 
\begin{equation} 
\Gamma_k= \frac{(N_k-I_k-R_k) I_k}{N_k}\simeq I_k,
\end{equation}
where we have neglected all terms of order $I_k^2$ and considered that
$R_k$ is of the same order of $I_k$ in the early stage of the
dynamics.
The simplification of the reaction kernel allows to write
 eqs.~(\ref{eq:unif_diff}) for the evolution of the density of infectious
individuals in the following form:
\begin{eqnarray}
\pt I_k= -p_k I_k + (1-p_k)(\beta-\mu)I_k+~~~~~~~~~\nonumber\\
~~~~~~~~~+\frac{k}{\langle k\rangle} \sum_{k'}k'P(k')d_{k'k}\left[
(1-\mu+\beta)I_{k'}\right].
\end{eqnarray}
Explicit solutions to the above set of equations for the early dynamics
of infectious individuals in degree block $k$ can be found
by considering the specific diffusion processes already introduced in the
previous section.

%*****************************************************************************
\Subsection{Traffic dependent mobility rates}\label{subsec:epi_traffic_diff}
By considering a uniform $p$ and the expression for $d_{kk'}$ 
of eq.~(\ref{hetcoeff}) we obtain after
some simple algebra the following dynamical reaction-rate equations
\begin{equation}
\pt I_k= -p I_k + (1-p)(\beta-\mu)I_k +p \frac{k^{(1+\theta)}}{\langle k^{(1+\theta)}
\rangle}\left[(1-\mu +\beta)\bar{I}\right],
\label{solutionI}
\end{equation}
that depend only on the densities of infectious individuals, and where
$\bar{I}=\sum_{k'}P(k')I_{k'}$.
A solution for the global density of infectious individual in the
metapopulation system is obtained by averaging both terms of the
equation over $P(k)$, obtaining:
\begin{equation}
\pt \sum_k P(k) I_k=\pt \bar{I}= (\beta-\mu)\bar{I},
\end{equation}
where we have considered that $\sum_k P(k)k^{(1+\theta)}= \langle k^{(1+\theta)}
\rangle$. This equation has the simple solution
\begin{equation}
\bar{I}= \bar{I}(0) e^{(\beta-\mu)t},
\end{equation}
where $\bar{I}(0)$ is the initial number of infected individuals in the
metapopulation system. It readily states that the overall
density of infectious individuals in the system can grow only if
$\beta>\mu$, thus recovering the epidemic threshold condition $R_0=\beta/\mu>1$.
The metapopulation system exhibits at the deterministic level an epidemic
threshold condition equivalent to that of each single
population that sets the time scale for the whole system. 
Intuitively this is stating that if the epidemic is not
able to proliferate in each local subpopulation, then it  cannot
produce a major outbreak at the metapopulation level. 

It is possible to solve the
early time behavior for all subpopulations of a given degree block
$I_k(t)$ by plugging in the explicit solution of $\bar{I}(t)$ in
eqs.~(\ref{solutionI}). This yields:
\begin{equation}
I_k(t)=A\frac{k^{1+\theta}}{\langle k^{1+\theta} \rangle} e^{(\beta-\mu)t}+
C_k e^{\left[ (1-p)(\beta-\mu)-p \right]t}
\end{equation}
where $A$ and $C_k$ are parameters fixed by the initial conditions. 
If we assume that
the metapopulation system is seeded with a total number of $I_0$ 
initially infected individuals
distributed homogeneously among subpopulations, i.e. 
$I_k(0)=I_0/V=\bar{I}(0)\,\,\,\forall k$, 
we obtain:
\begin{equation}
A=\bar{I}(0) \quad \textrm{and}\quad  C_k=\bar{I}(0) 
\left(1-\frac{k^{1+\theta}}{\langle k^{1+\theta} \rangle}  \right)
\label{eq:epik1}
\end{equation}
If the $I_0$ infected are distributed only in the $k_0-$block subpopulations,
$I_k(0)=\delta_{k,k_0}\bar{I}(0)/P(k_0)$, the coefficients $A$ and $C_k$
assume the following values:
\begin{equation}
A=\bar{I}(0) \quad \textrm{and}\quad  C_k=\bar{I}(0) 
\left(\frac{\delta_{k,k0}}{P(k_0)}-\frac{k^{1+\theta}}
{\langle k^{1+\theta} \rangle}  \right)\,.
\label{eq:epik2}
\end{equation}
The above results show how the choice of initial conditions, 
whether if homogeneously
distributed or locally distributed, affects the early stages of the epidemic
outbreak inside subpopulations of different block $k$. The change in
the initial stage of the disease evolution in subpopulations 
depending on the degree block is confirmed by the numerical results
reported in section 6.

%*****************************************************************************
\Subsection{Population dependent diffusion rate}
\label{hetrate}
If we consider a population dependent diffusion rate, $p_k=T_k/N_k$,
the system's behavior will be given by plugging into 
the set of eqs.~(\ref{eq:unif_diff}) the degree dependent diffusion
probability $p_k$ and the expression of the rate
of diffusion on a link $k' \to k$, 
$d_{k'k} = \frac{w_0(k'k)^{\theta}}{N_{k'}}$.
In the approximation of early stage dynamics and considering the normalization condition 
$T_k=k^{1+\theta}  w_0 \langle k^{1+\theta}\rangle/\langle k\rangle$, 
we obtain:
\begin{eqnarray}
\pt I_k=-p_kI_k+(1-p_k)(\beta-\mu)I_k+\nonumber\\
+\frac{k^{1+\theta}}{\langle k^{1+\theta}\rangle}(1+\beta-\mu)\Omega,
\end{eqnarray}
where $\Omega=\sum_k P(k)p_k I_k$. Proceeding along the line followed
in the previous section, the solution can be found for the early stage
behavior of the average number of infectious individuals $\bar{I}=\sum
P(k)I_k$ by averaging both terms of the equation over $P(k)$:
\begin{equation}
\pt \bar{I} = (\beta-\mu) \bar{I}\,,
\end{equation}
yielding
\begin{equation}
\bar{I}=\bar{I}(0)e^{(\beta-\mu)t}
\end{equation}
and thus recovering the epidemic threshold 
condition $R_0=\beta/\mu>1$ also
in this case. The early stage behavior of the epidemic in the 
deterministic approximation for the whole system does not differ from
what observed in the previous heterogeneous frame.
It is worth to mention that the solution for the population
dependent diffusion rate is obtained under
very general conditions for the size of the subpopulations $N_k$ which
can assume any value subject only to the constraint $N_k>T_k$ to ensure a 
proper definition of the probability of diffusion.

It is clear from the previous analysis that the deterministic equations
are not capable to account for the invasion threshold as they consider
the diffusion and reaction
processes in a mean-field perspective that provides deterministic
equations for the average values. Indeed, the deterministic diffusion process
allows any fraction $pI_k$ of infected 
individual to deterministically seed new populations, washing out the
stochastic effects responsible for the invasion threshold. While these
equations do not allow to capture the invasion threshold, they provide
a good description of the system and its behavior across degree
classes above the invasion threshold, as we will show in the next
section by comparing the analytical results with stochastic
simulations at the mechanistic level. 

%*******************************************************************
%*******************************************************************

\Section{Mechanistic numerical simulations}
\label{simulations}
Here we provide extensive numerical simulations to support the
theoretical picture described above. We report results from Monte Carlo
simulations in a variety of different cases and compare them with the 
analytical findings. We adopt mechanistic numerical simulations where each
single individual is tracked in time, during both the infection dynamics 
and the
diffusion processes.  The system evolves following a stochastic
microscopic dynamics and at each time step it is possible to monitor
quantities which depend on the subpopulation~-~such as e.g.  
the number $I_j(t)$ of infectious individuals in the subpopulation $j$ at 
time $t$~-~and
also averages over blocks of nodes~-~e.g. the average number
$I_k(t)$ of infected in subpopulations with degree $k$~-~or over the whole
system, $\bar{I}(t)$. In addition, it is possible to study the evolution
of the epidemic by monitoring the invasion dynamics at the local population
level, and therefore measure the number of diseased subpopulations 
at time $t$, $D(t)$.  
Given the stochastic nature of the dynamics,
the experiment can be repeated with different realizations of the
noise, different underlying graphs, and different initial conditions.
This approach is equivalent to the real evolution of epidemic processes in
the generated networks and can be used to validate the theoretical
results obtained in the analytical approach.

The substrate network is given by an uncorrelated complex network generated
with the uncorrelated configuration
  model~\citep{Catanzaro:2005}, based on the
\emph{Molloy-Reed}  algorithm~\citep{MR} with an additional
constraint on the possible maximum value of the degree in order to
avoid inherent structural correlations.
The algorithm is defined as
follows.  Each node $i$ is assigned a degree $k_i$ obtained from a
given degree sequence $P(k)$  subject to the restriction $k_i<V^{1/2}$.
Here we assume a power-law degree distribution, $P(k)\sim k^{-\gamma}$,
 with $\gamma=2.1$ and $\gamma=3$.
Links are then drawn to randomly connect pairs of nodes, respecting
their degree and avoiding self-loops and multiple edges.  Sizes of
$V=10^4$ and $V=10^5$ nodes have been considered.  
Weights on the connections among subpopulations are defined 
following the statistical law found in real transportation system
(see eq.~(\ref{eq:w_kk'})). Therefore the weight on the link
between subpopulation $i$ and subpopulation $j$ is given by:
\begin{equation}
w_{ij}=w_0 (k_i k_j)^{\theta},
\end{equation}
where $k_i$ and $k_j$ are the degrees of the subpopulations $i$ and $j$, respectively.
Here we fix $w_0=1$, whereas $\theta$ assumes different values, including 
$\theta=0$ for uniform weights. This expression for the weights
is then used to define the diffusion rates in the cases analytically
investigated, as in eq.~(\ref{hetcoeff}) and eq.~(\ref{eq:pop_coeff}).

The dynamics proceeds in parallel and considers
discrete time steps representing the unitary time scale $\tau$ of the
process. The reaction and diffusion rates are therefore converted into
probabilities and at each time step, the system is updated according
to the following rules.  a) Infection dynamics: \emph{i)} The 
contagion process assumes that in each subpopulation $j$
individuals homogeneously mix and have a finite number of contacts, 
so that the probability for a susceptible to contract
a virus from an infected is proportional to the transmission
rate and normalized to the subpopulation size, $\beta/N_j$. 
At each time step in the simulation, each susceptible is turned into
an infectious with probability $1-(1-\frac{\beta}{N_j}\tau)^{I_j}$.
 \emph{ii)} At the same time, each infectious individual is subject to the
recovery process and becomes recovered with probability $\mu\tau$.
b) After all nodes have been updated for the
reaction, we simulate the diffusion process. Results shown in the
following subsections refer to the traffic dependent diffusion rate.

%*****************************************************************************

\Subsection{Global and local threshold in heterogeneous metapopulation models}

In the previous sections we have shown that along with the usual
local epidemic threshold $R_0>1$, 
the stochasticity and discreteness of the metapopulation diffusion
process induce an intrinsic 
invasion threshold $R_*>1$~-~at the global level~-~which
rules the invasion dynamics in the coarse-grained view of the 
system. This threshold determines
whether the coupling between subpopulation is high enough in order to allow
the virus to spread from one subpopulation to another and invade a finite portion
of the whole system.
Here we numerically investigate this phenomenon by studying a metapopulation
model with heterogeneous structure ($P(k)\simeq k^{-\gamma}$) and varying
the coupling force between subpopulations. Initially, let us consider 
that the diffusion probability of an individual on the
heterogeneous metapopulation structure is locally independent of the degree
of the subpopulation~-~i.e. $p_k=p$~-~and heterogeneous on the links
departing from a given subpopulation, following eq.~(\ref{hetcoeff}).
The probability of diffusion from a subpopulation $i$ to a subpopulation $j$
for each individual in any given compartment located in $i$ is therefore given by:
\begin{equation}
d_{ij}=p\frac{w_0 (k_i k_j)^{\theta}}{T_i},
\end{equation}
where $T_i=\sum_j w_{ij}$ represents the traffic in $i$.
The simulations proceeds according to the following procedure: 
at each time step, after the update
for the local infection dynamics within the subpopulations (see
previous section),
each individual in any compartment in subpopulation $i$ moves to a neighboring 
subpopulation $j$ with probability $d_{ij}$. 
We analyze different values
of $R_0$ by assuming $\mu=0.2$ and $\mu=0.02$, and varying the value of the transmission
rate $\beta$. The simulations start with a localized initial condition
given by the seeding of a subpopulation having degree $k_0$ with $I_0=10$
infected individuals. This allows to monitor of the epidemic evolution
in the metapopulation model and measure the final size of the
epidemic, expressed in terms of the number or density of cases obtained in the
whole system and the number of subpopulations experiencing an outbreak. 

\begin{figurehere}
\begin{center}
\includegraphics[width=8.5cm]{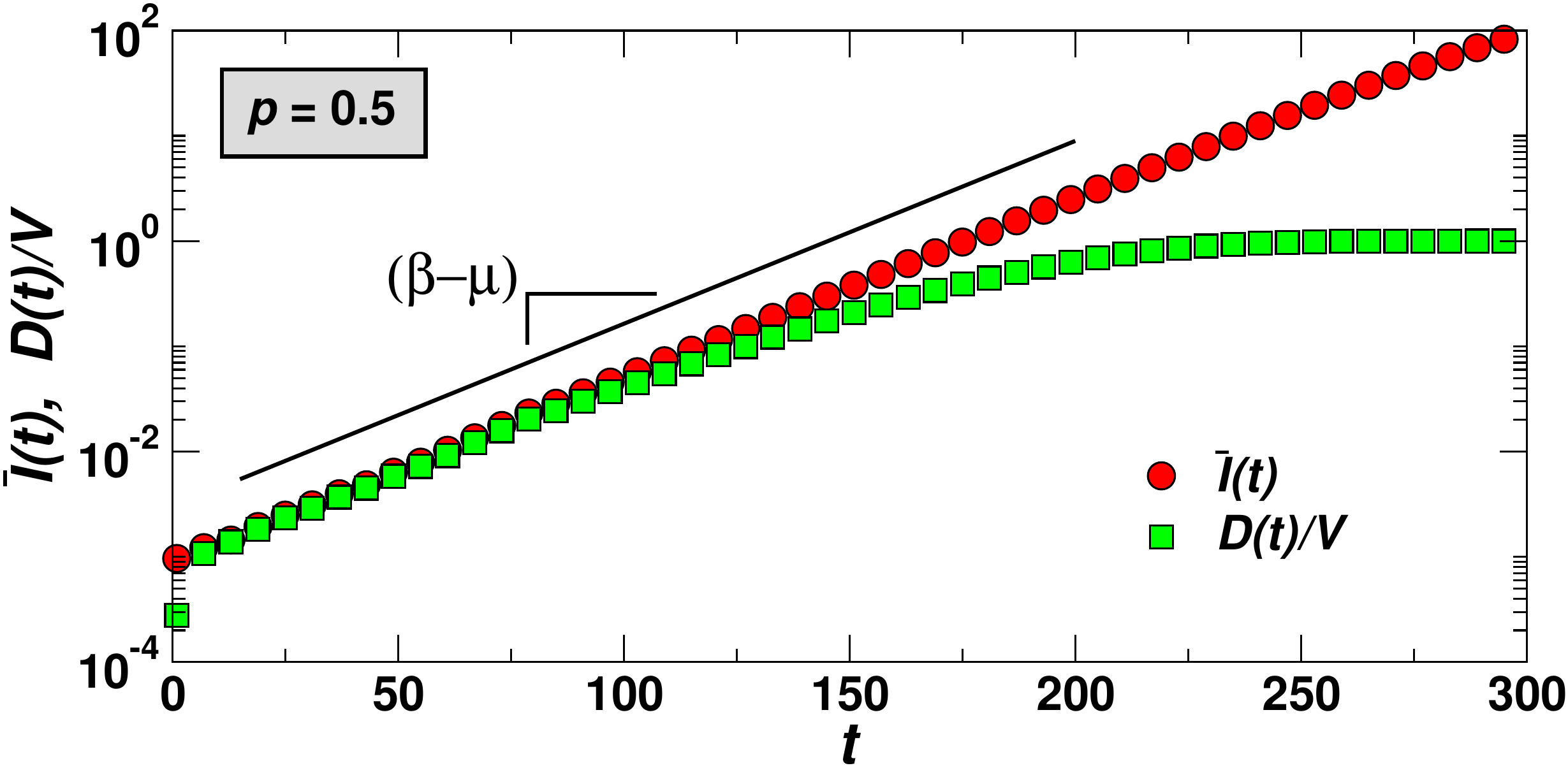}

\includegraphics[width=8.5cm]{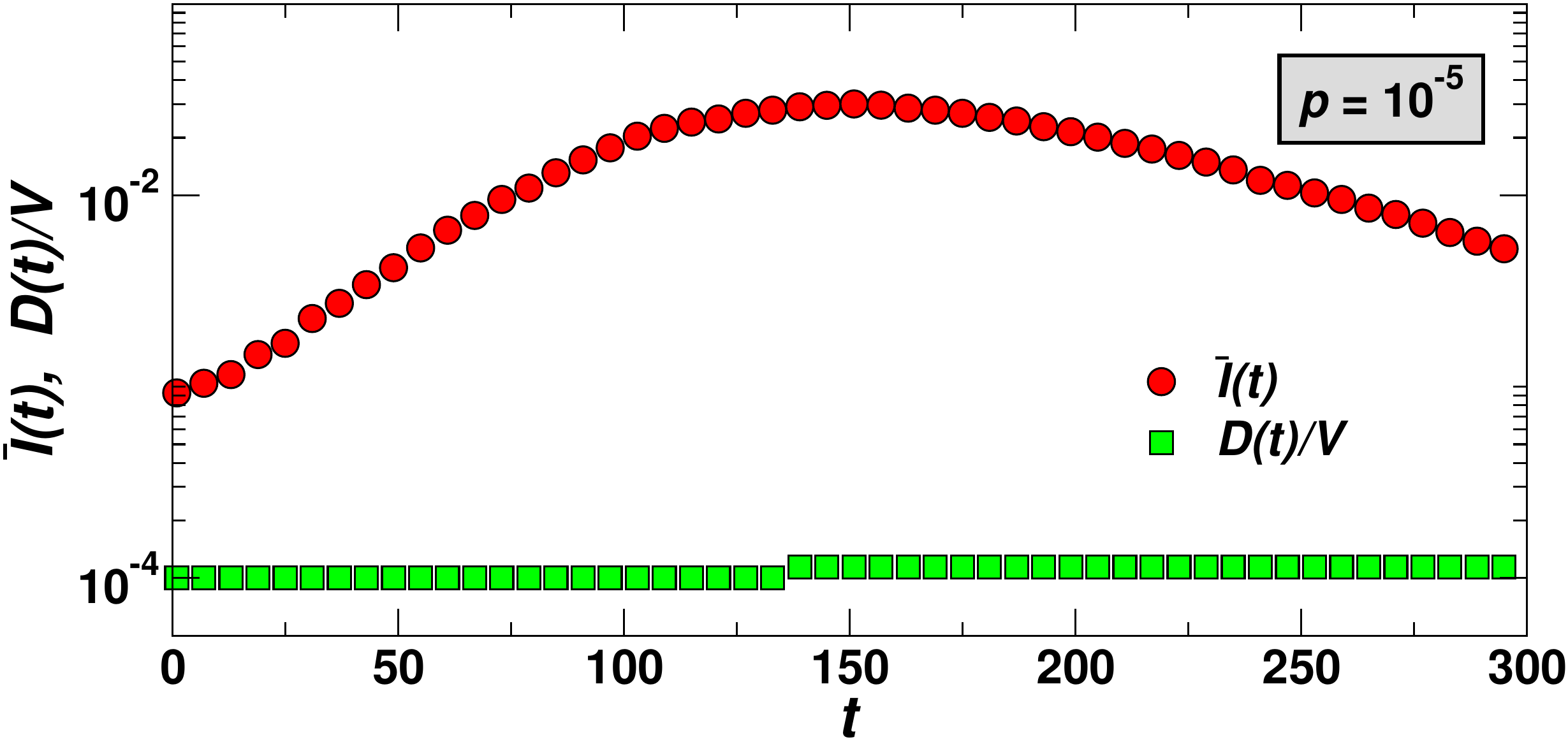}
\renewcommand{\baselinestretch}{1.0}
\caption{\small Metapopulation system's behavior above and below the
global threshold. Results refer to $R_0=3$, $\bar{N}=10^3$, and
$\theta=0.5$. The epidemic is in both cases above the local
threshold, leading to an exponential increase 
of $\bar{I}(t)$. Differences in the diffusion probability values (top: $p=0.5$,
bottom: $p=10^{-5}$) show the effect of the global threshold on the number $D(t)$
of diseased subpopulations. $D(t)$ is normalized to the system size $V=10^4$
for sake of visualization.}
\label{fig:double-th}
\end{center}
\end{figurehere}

In Figure~\ref{fig:double-th}  
we analyze the behavior of $I(t)$ and $D(t)$ for $R_0=3$, above 
the local threshold, and for two values of the diffusion rate $p=0.5$
and $p=10^{-5}$ that poise the system below and above the invasion
threshold, respectively. While in both cases the figure shows an
increase of the value of $I(t)$, the behavior of $D(t)$ is very
different above and below the invasion threshold. Indeed, above the
invasion threshold the number of affected subpopulations is increasing
exponentially, while below the threshold the number of subpopulations
remains small and goes to zero in a finite time. The increase in
$I(t)$ instead is guaranteed also below the invasion threshold by the
outbreak in the initial seeded population. On a longer time however
$I(t)$ keeps increasing only if the system is above the invasion
threshold and new subpopulations are progressively infected. 

\end{multicols}

\begin{figurehere}
\begin{center}
\includegraphics[width=0.9\textwidth]{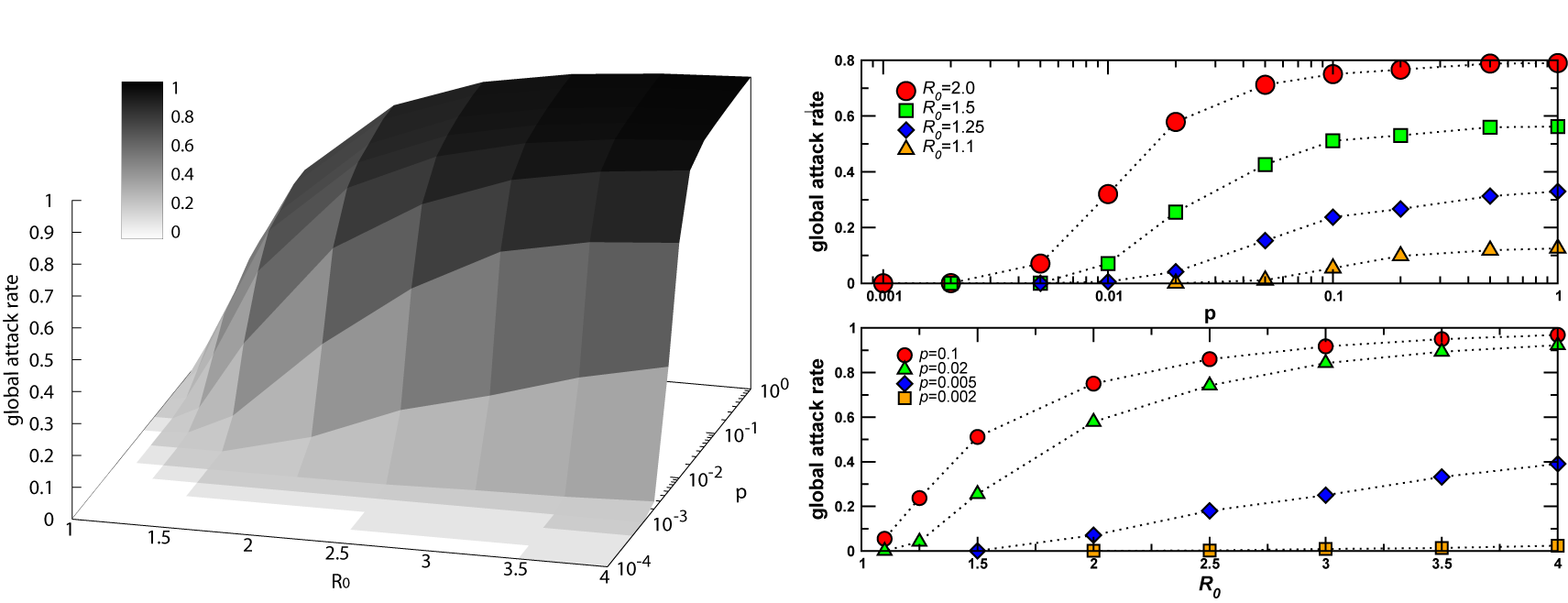}
\renewcommand{\baselinestretch}{1.0}
\caption{\small Global threshold in a  heterogeneous metapopulation
  system with traffic dependent diffusion rates. 
The left panel shows a 3D surface representing the
value of the final epidemic size in the metapopulation system as a function of 
the local threshold $R_0$ and of the diffusion probability $p$. If $R_0$
approaches the threshold, larger values of the diffusion probability $p$
need to be considered in order to observe a global outbreak in the
  metapopulation system. On the right, two plots showing the cross
  sections of the 3D plot 
at fixed values of $R_0$ (top) and at fixed values of the traveling rate $p$ 
(bottom).}
\label{fig:glth}
\end{center}
\end{figurehere}

\begin{multicols}{2}

While Figure~\ref{fig:double-th} provides a clear evidence of the two
separate threshold mechanisms, a complete analysis of the system phase
diagram is obtained by analyzing the behavior of the global
metapopulation attack rate, defined as the total fraction of cases
$R(\infty)/N$ at the end of the epidemic, as a function of
both $R_0$ and $p$. In Figure~\ref{fig:glth}, we report the 
global attack rate surface in the $p$-$R_0$ space, and the
two dimensional plots of the $p$ and $R_0$ crosscuts.  
Figure~\ref{fig:glth} clearly shows the effect of different couplings
as expressed by the value of $p$ in reducing the final size of the epidemic
at a given fixed value of $R_0$. The smaller the value of $R_0$, the higher 
the coupling needs to be in order for the virus to successfully invade
a finite fraction of the subpopulations, in agreement with the
analytic result of eq.~(\ref{eq:glth1}). This provides a clear
illustration of the varying global invasion threshold as a function of
the reproductive rate $R_0$. On the contrary, $p$-crosscuts show that
whatever the value of $p$, $R_0<1$ does not allow the epidemic to
spread.  
\begin{figurehere}
\begin{center}
\includegraphics[width=8.5cm]{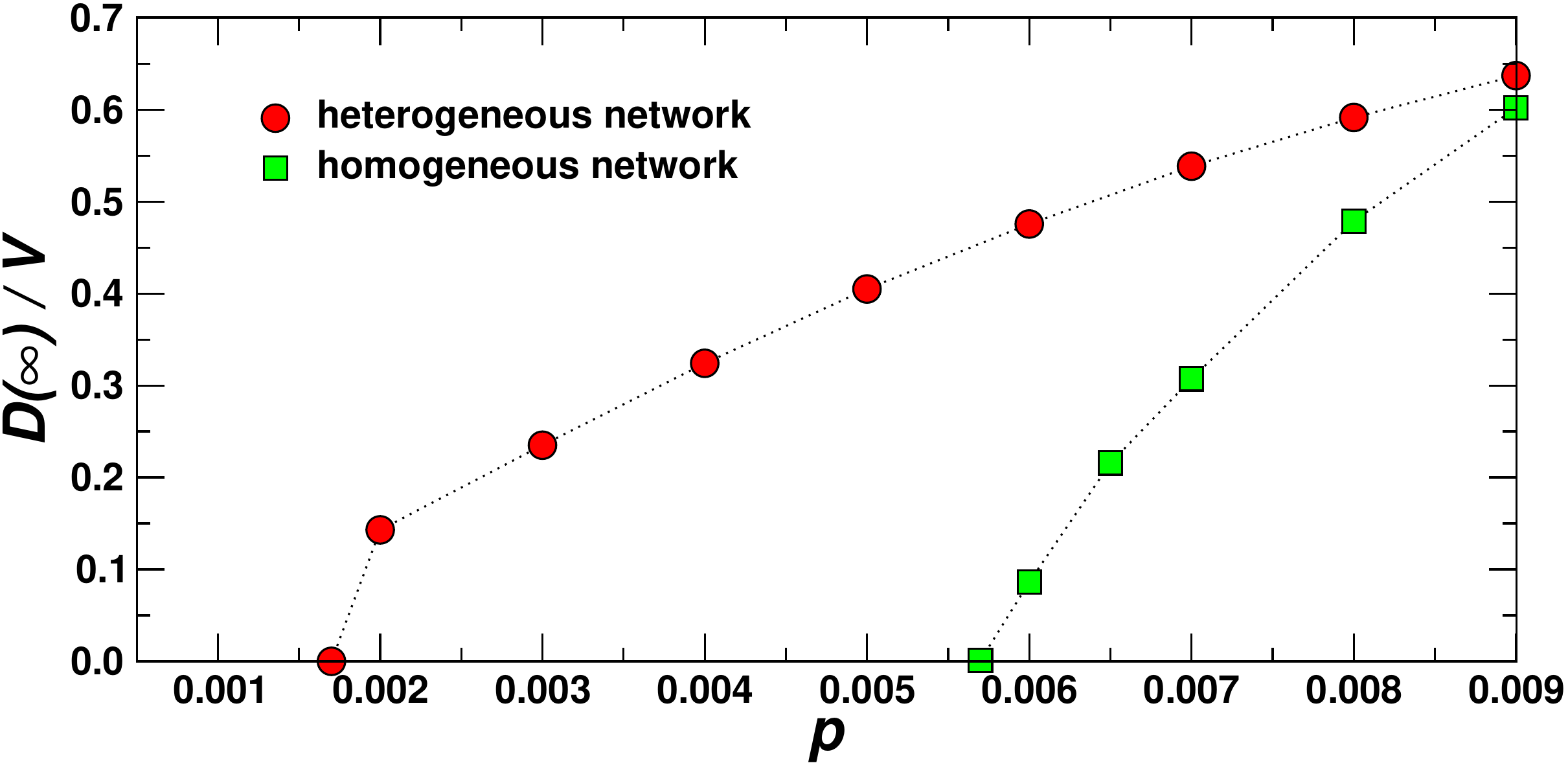}

\renewcommand{\baselinestretch}{1.0}
\caption{\small Effect of metapopulation structure heterogeneity  on 
the global epidemic threshold. The final fraction of diseased 
subpopulations $D(\infty)/V$ at the end of the global epidemic 
is shown as a function of the traveling diffusion rate $p$.
A heterogeneous network with heavy-tailed degree distribution, $P(k)\sim k^{-2.1}$ is compared to a homogeneous network with poissonian $P(k)$
having the same size $V=10^5$ and
same average degree. Here $\theta=0$.}
\label{fig:glth2}
\end{center}
\end{figurehere}
Finally, it is possible to study the effect of the heterogeneity
of the metapopulation structure. Figure~\ref{fig:glth2}
shows the results obtained comparing a heterogeneous network
characterized by a scale-free degree distribution $P(k)\sim k^{-2.1}$
with a homogeneous network having the same size $V=10^5$ and
same average degree.
The presence of topological fluctuations lead to a smaller ratio
$\langle
k^{1+\theta} \rangle^2/(\langle k^{2+2\theta}\rangle-\langle k^{1+2\theta}\rangle)$, thus
lowering the value of the mobility threshold with respect to the 
homogeneous network.

%*****************************************************************************
\Subsection{Epidemics above the invasion threshold}

Above the global invasion threshold $R_*>1$, the epidemic process is
guaranteed to invade a macroscopic fraction of subpopulations and it
is possible to inspect the validity of 
the results obtained in Section 4 with the deterministic 
reaction-diffusion equations. A general conclusion is that 
the global density of infectious individuals in the system
in the early stage of the epidemic dynamics grows as
\begin{equation}
\bar{I}(t)=\bar{I}(0)e^{(\beta-\mu)t},
\label{eq:avI_traffic_diff}
\end{equation}
if the threshold condition, $R_0=\beta/\mu>1$ is satisfied. 
The early time 
behavior expressed in the above equation is also independent 
of the parameters related to the diffusion process among subpopulations,
such as the homogeneous diffusion rate $p$ and the exponent
$\theta$ which governs the relation between weights and subpopulation degrees.
The analytic result of eq.~(\ref{eq:avI_traffic_diff}) is confirmed in 
Fig.~\ref{fig:epi_traffic_diff}, where we show simulation results of the
metapopulation epidemic model with traffic dependent diffusion rates.
We consider systems of $V=10^4$ subpopulations each of initial size $\bar{N}=10^4$,
connected through a heterogeneous
network having degree distribution $P(k)\simeq k^{-3}$. The disease parameters
assume the following values: $\beta=0.04$ and $\mu=0.02$, yielding $R_0>1$.
The simulations are seeded with $\bar{I}(0)=100$ infectious individuals,
homogeneously distributed among  subpopulations. 
Both homogeneous ($\theta=0$) and heterogeneous ($\theta=0.5$)
diffusions are considered, as well as different values of the diffusion
probability $p=0.5,\,0.75,\,1.0$.

Results in Fig.~\ref{fig:epi_traffic_diff} show that the early behavior of the
global density of infectious individuals is independent of the values of $\theta$
and $p$, and of the location of the initial seed, whether if homogeneously
distributed among the subpopulations of a given degree block or
localized 
in a single subpopulation. All simulations show an exponential
increase 
which confirms the analytic findings.

\begin{figurehere}
\begin{center}
\includegraphics[width=8.5cm]{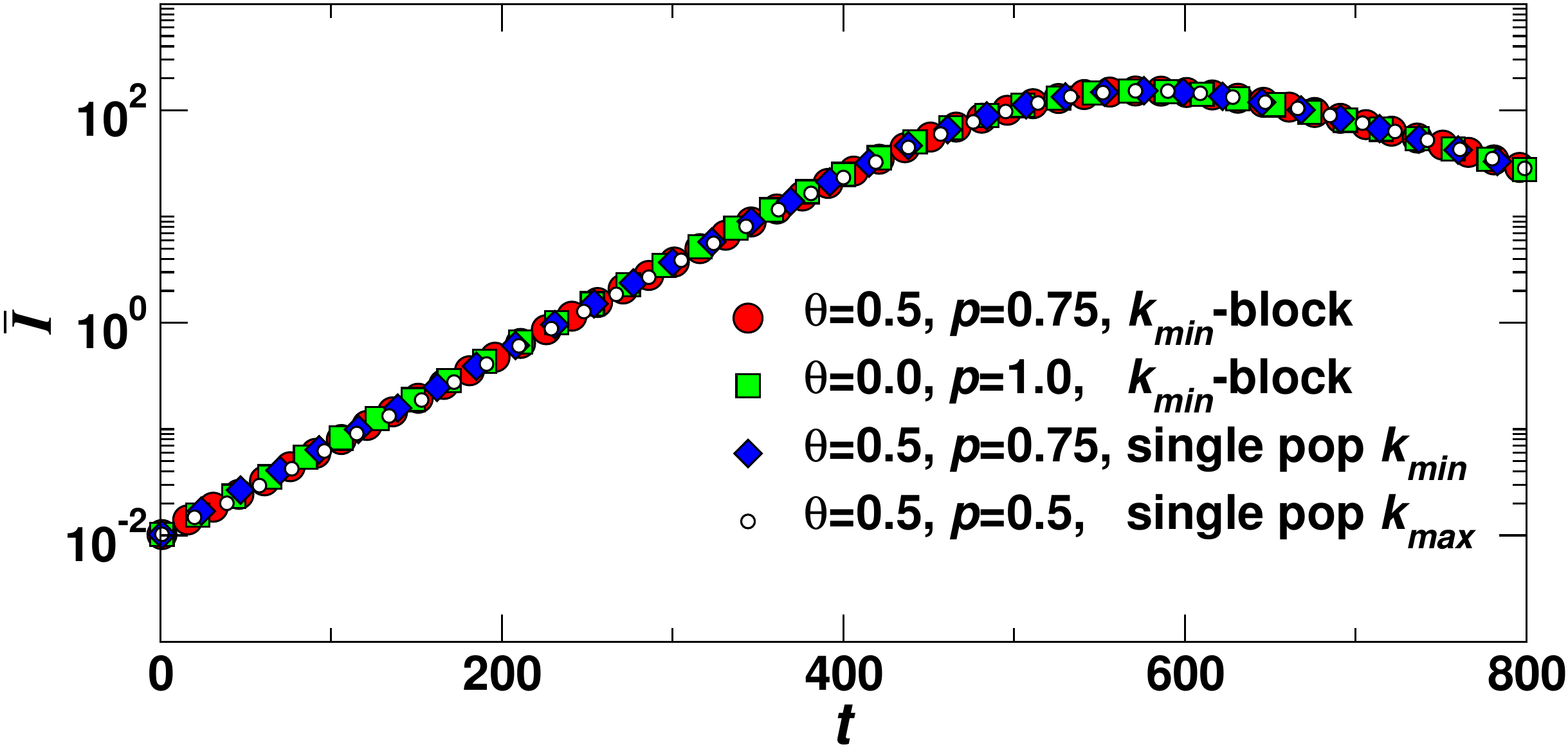}

\includegraphics[width=8.5cm]{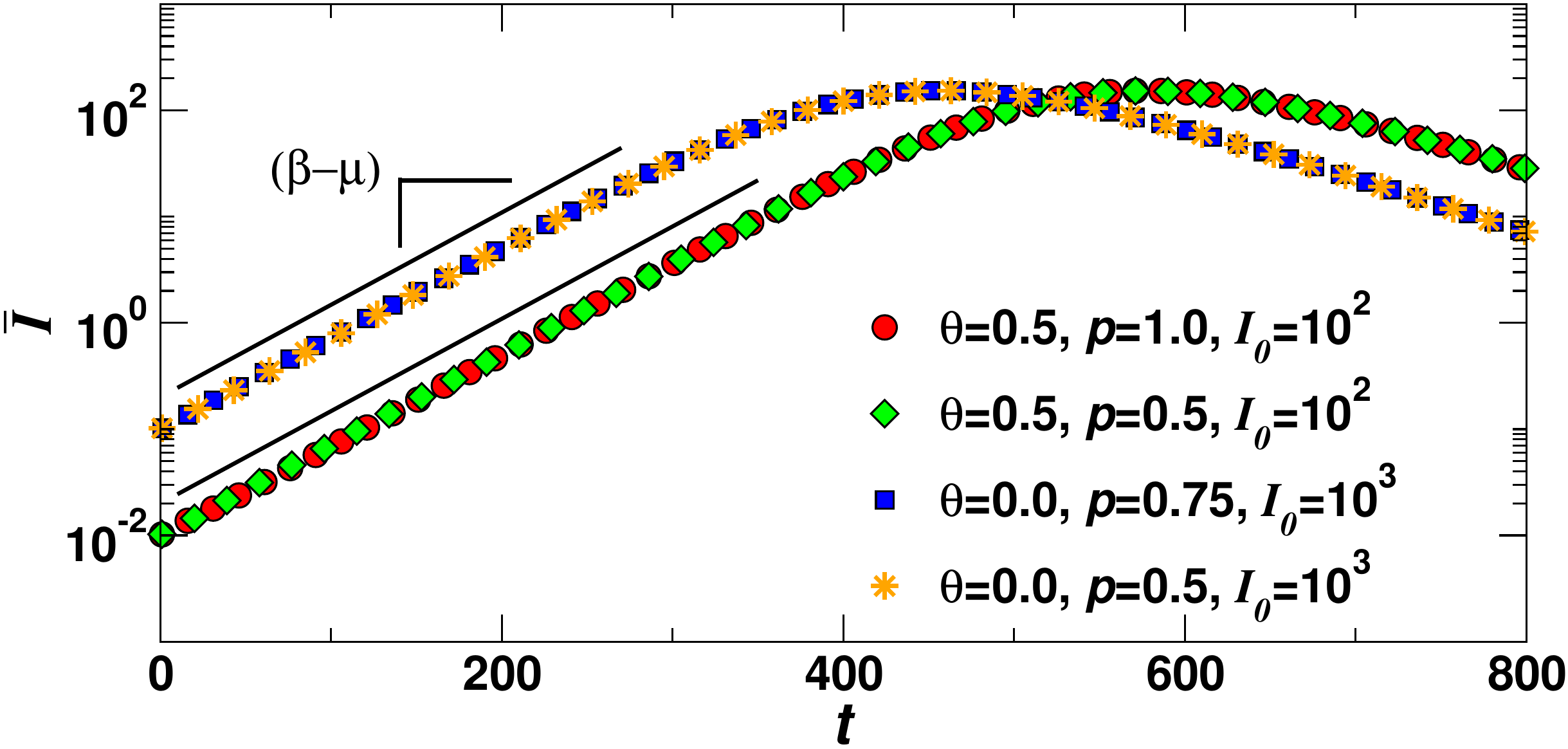}
\renewcommand{\baselinestretch}{1.0}
\caption{\small Evolution in time of the global density of infectious individuals
in a  heterogeneous metapopulation system with traffic dependent diffusion rates.
Top: the effects of the heterogeneity of diffusion ($\theta$), of the probability
of diffusion ($p$), and of the distribution of initially infected individuals
in the system (homogeneously distributed in a $k-$degree block or concentrated in
a single subpopulation) are compared and found
to produce the same early stage behavior. Bottom: changes in the initial condition 
value $\bar{I}$ produce the same exponential increase in the metapopulation
system behavior.}
\label{fig:epi_traffic_diff}
\end{center}
\end{figurehere}

Numerical simulations also allow for the study of the dynamic behavior
of infectious individuals in subpopulations of degree block $k$. The solutions
obtained in subsection 4.4 show a dependence
of the early time behavior on the degree $k$ of the subpopulation, pointing
to a dynamics which switches on degree modes at different times. Results
of the numerical simulations confirm this findings, as shown 
in Fig.~\ref{fig:epik_traffic_diff}. Here the disease 
parameters assume the same values as before,
 and the diffusion is governed by the values $\theta=0.5$ and $p=0.75$ for the 
numerical results reported in the top
panel, whereas the effect of different values of $\theta$  is reported in
the bottom panel.
In order to see the effects of different initial conditions on the dynamical
behavior of degree block $k$ subpopulations 
(see eqs.~(\ref{eq:epik1}) and (\ref{eq:epik2})),
we seed the epidemics with \emph{i)} $10^2$ infectious individuals homogeneously
distributed among subpopulations, or with \emph{ii)} $10^2$ infectious individuals
localized in subpopulations of degree block $k_0$ (results in the top panel
correspond to $k_0=k_{max}$). While the global behavior
$\bar{I}(t)$ is not affected by the choice of the initial conditions (see
previous Figure), the subpopulations experience outbreaks at different times,
as brought and delayed by the diffusion dynamics. The system heterogeneity,
as contained in the factor $k^{1+\theta}/\langle k^{1+\theta} \rangle$ of the
explicit solution of $I_k(t)$, differentiates the evolution of the degree block
subpopulations at short times. Numerical results obtained for the study of the 
effect of traffic heterogeneity (Fig.~\ref{fig:epik_traffic_diff} bottom)
are compared with the analytical findings of subsection 4.4.

\begin{figurehere}
\begin{center}
\includegraphics[width=8.5cm]{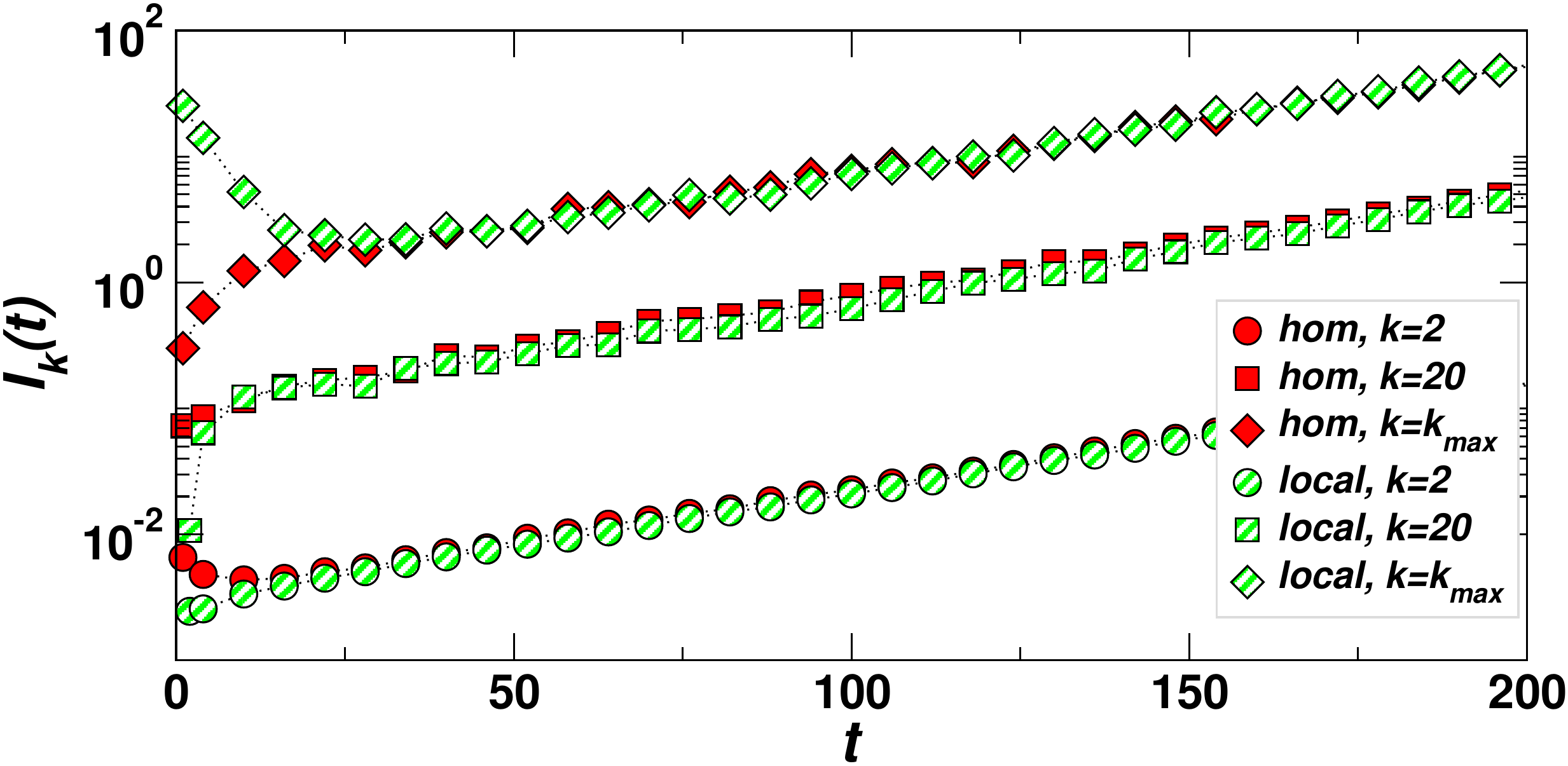}

\includegraphics[width=8.5cm]{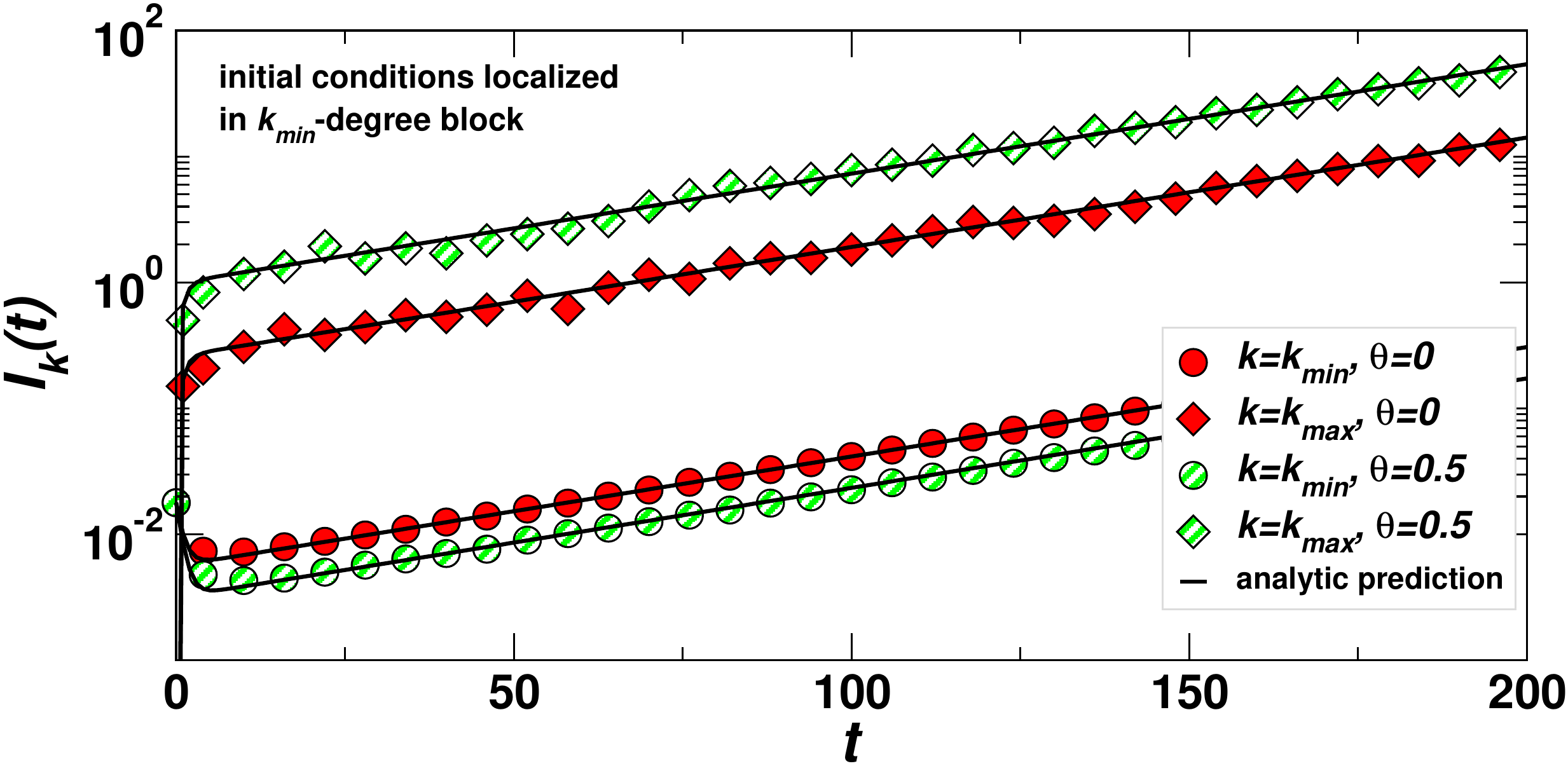}
\renewcommand{\baselinestretch}{1.0}
\caption{\small Evolution in time of the  density of infectious individuals
in degree block $k$ subpopulations.
Top: changes in the initial conditions (homogeneous \emph{vs.} localized
in $k_{max}$)
yield different behaviors in the early time dynamics of distinct degree blocks. 
Bottom: changes in the value of $\theta$ impact differently the time evolution
of the average number of infected $I_k$ in each degree block, whether if
$k=k_0$ (i.e. $k_{min}$) or $k\ne k_0$ (i.e. $k=k_{max}$).}
\label{fig:epik_traffic_diff}
\end{center}
\end{figurehere}

%*****************************************************************************
%*****************************************************************************
\Section{Conclusions and outlook}
Here we have introduced an analytic framework in terms
of degree block variables which allows to gain insights on the behavior
of mechanistic metapopulation epidemic models which explicitly include
demographic and mobility heterogeneities. The system is shown to display
a local epidemic threshold which depends on the disease parameter values only and
is responsible for the epidemic outbreak at the local scale, and a
global epidemic threshold which determines the invasion dynamics of the 
subpopulations and depends critically on the disease parameters and the 
diffusion rates
of the individuals. Changes in coupling between subpopulations are shown to
have critical implications for disease extinction.

The results provide useful insights for the basic theoretical understanding
of mechanistic epidemic models in complex environments, which can then be used to
build more realistic data-driven large-scale computational approaches for
real case scenarios and spatially targeted control measures.
However, several key theoretical and practical issues are still to be addressed.
Data on human dynamics at the local level, i.e. within any subpopulation,
could push forward a more sophisticate theoretical framework for the
local infection dynamics, to go beyond the homogeneous mixing assumption~\citep{Meyers:2005,Lloyd-Smith:2005}.
The behavior of metapopulation models characterized by complex internal
structure in each patch is a major question for the theoretical epidemiology
of the future. In addition, more realistic and detailed diffusion patterns
should be included in order to better model the coupling terms by including
non-Markov processes and introducing elements of memory in the system.
Obviously this corresponds to the need for more accurate data on population
behavior, such as fraction of commuters, probability of short/medium/long range 
travel, trip duration, and so on~\citep{Riley:2007}. Additional levels of heterogeneity
can be also included in the diffusive patterns by introducing a dependence
of the probability of diffusion
on the stage of the disease. In many real cases, e.g. the severity of symptoms
or hospitalization measures would prevent the diffusion out of a patch 
to a portion of the population. The impact of heterogeneities in traveling pattern
of individuals depending on their infection state could provide additional
insights fundamental to the study of global extinction and eradication.
All these improvements and future directions would help filling the
gap between the evidence from increasingly realistic epidemic models 
and their theoretical understanding.

%%%%%%%%%%%%%%%%%%%%%%%%%%%%%%%%%%%%%%%%%%%%%%%%%%%%%%

\vspace*{0.5cm}

{\small
We are grateful to Alain Barrat, Marc Barthelemy and Romualdo Pastor-Satorras for 
useful discussions during all stages of preparation of this work. 
A.V. is partially funded by the NSF award IIS-0513650. V.C and
 A.V. are partially funded by the the CRT foundation through the
 Lagrange Project.}

\vspace*{0.5cm}

{\small

\renewcommand{\baselinestretch}{0.9}

}

\end{multicols}

\end{document}